\documentclass[10pt,conference,letterpaper]{IEEEtran}
\usepackage{amsmath,amssymb,amsthm}
\usepackage{enumerate}
\usepackage{stfloats}
\usepackage{comment}

\usepackage[T1]{fontenc}
\usepackage{graphics} 
\usepackage{epsfig} 
\usepackage[mathscr]{euscript}
\usepackage{algorithm}
\usepackage{algorithmic}

\usepackage{tikz}
\usetikzlibrary{arrows,shapes,chains,matrix,positioning,scopes,patterns,calc}
\usepackage{pgfplots}
\pgfplotsset{compat=1.3}
\usepgflibrary{shapes}

\renewcommand{\epsilon}{\varepsilon}

\newcommand{\RNum}[1]{\uppercase\expandafter{\romannumeral #1\relax}}

\DeclareMathAlphabet{\mcl}{OMS}{cmsy}{m}{n}

\newlength\tikzwidth
\newlength\tikzheight

\def\avg{\ensuremath{\mathrm{avg}}}

\def\RS{\ensuremath{\mathrm{RS}}}
\def\PA{\ensuremath{\mathrm{PA}}}
\def\IRSA{\ensuremath{\mathrm{IRSA}}}

\def\rf{\ensuremath{\mathrm{ref}}}
\def\tot{\ensuremath{\mathrm{tot}}}

\begin{document}
\title{{Rate Selection and Power Adaptation\\
using Maximal Ratio Combining\\
for the Random Access Gaussian Channel}}
\author{Arman Hasanzadeh, Jean-Francois Chamberland, Krishna Narayanan \\
Department of Electrical and Computer Engineering, Texas A\&M University\\
{\tt\small {\{armanihm,chmbrlnd,krn\}@tamu.edu} }
}

\maketitle

\begin{abstract}
With the emergence of machine-driven communication, there is a renewed interest in the design of random multiple access schemes for networks with large number of active devices.
Many of the recently proposed access paradigms are enhancements to slotted ALOHA.
One of the popular schemes, irregular repetition slotted ALOHA (IRSA), is based on an analogy between multiple access with successive interference cancellation and message-passing decoding on bipartite graphs.
Most of the results on IRSA and its variants focus on the collision channel and they ignore physical limitations such as transmit power constraints and additive Gaussian noise at the physical layer.
As such, naive extensions of IRSA to the Gaussian multiple access channel are not power efficient in the low signal-to-noise-ratio regime.
This work introduces a novel paradigm whereby devices adapt their rates
and/or transmit powers
based on their chosen repetition profiles.
The receiver performs maximal ratio combining over all the replicas prior to decoding a message.
Numerical results for finite number of users show that the proposed scheme can provide substantial improvements in terms of power efficiency and average rate over standard IRSA.
\end{abstract}


\section{Introduction}
\par Emerging wireless networks, such as the Internet of Things (IoT) and vehicular networks, are characterized by massive numbers of devices and unpredictable, bursty traffic.
In such networks, existing coordinated multiple access mechanisms are known to consume significant resources to facilitate coordination among active devices.
When the typical payloads of data packets are small, the overhead associated with coordination can result in a substantial throughput penalty.
This situation invites the examination of alternate frameworks.
In particular, efficient random multiple access schemes form a promising paradigm to manage networks with massive numbers of devices.

\par It is well known that traditional slotted ALOHA shows poor performance when the number of users is large; indeed, multiple access throughput is limited to $1/e \approx 0.37$~\cite{gallager1985perspective}.
Fortunately, important improvements to slotted ALOHA have been proposed recently.
A key enhancement is produced when collided packets are not dropped, but stored in a buffer and subsequently decoded via successive interference cancellation (SIC) which was first introduced in \cite{yu2007high} and later in \cite{gollakota2008zigzag}. Two important examples of such schemes are Contention Resolution Diversity Slotted ALOHA (CRDSA)~\cite{casini2007contention} and Irregular Repetition Slotted ALOHA (IRSA) \cite{liva2011graph,paolini2011high,paolini2015code}.
The latter article draws a pivotal connection between SIC and message-passing decoding on bipartite graphs in erasure channels.
This analogy has been leveraged to show that, in the absence of noise and without any power constraints at the transmitter, the soliton distribution is an optimal slot selection distribution in the sense that throughput can be arbitrary close to one when the number of users is asymptotically large~\cite{narayanan2012iterative}.

\par For additive Gaussian noise multiple access channels with power constraints, naive extensions of CRDSA or IRSA result in poor power efficiency, particularly in the low signal-to-noise ratio (SNR) regime.
This is mainly due to the fact that only one of the repetitions is used in the decoding process even though a packet may be repeated multiple times.
Thus, the energy transmitted in other slots is not harnessed effectively.
Madala et al. address this shortcoming in \cite{madalaicc2015} and propose a scheme where each user picks a rate randomly according to a carefully chosen probability distribution from a set of admissible rates.
It is shown that, asymptotically, the sum rate for all the users in a frame converges to the GMAC capacity with an additive gap of $\Theta(\log \log K)$ where $K$ is the number of users.
However, they consider a different framework in which there are no slots and all the users send at the same time. However, the number of distinct rates is equal to the number of users $K$ which makes the receiver very complex and impractical even for small number of users.

\par Kissling propose Contention Resolution ALOHA (CRA) as extensions of CRDSA to the asynchronous (unslotted) setting~\cite{kissling2011asynch}.
In CRA, users can have variable length packets and the repetition rate is fixed.
Clazzer et al.~\cite{clazzer2016exploiting,clazzer2017asynch} suggest using all replicas of a packet for decoding through maximal ratio combining (MRC) or selection combining (SC) to improve the performance of CRA in the presence of Gaussian noise.
We note that this modification can only improve performance.
Still, their algorithm, called Enhanced Contention Resolution ALOHA (ECRA), applies to the asynchronous (unslotted) setting and it is tailored to fixed repetition rates, like CRDSA.
Although their scheme shows good improvement over CRA, it is very sensitive to rate, i.e., for some rates, performance can be worse than CRDSA.
Moreover, the transmission scheme remains unaltered compared to CRDSA; only the operation of the receiver is altered, leaving room for further improvements.

\par In this article, we capitalize on this opportunity and seek to advance random multiple access schemes.
We consider a random multiple access channel with additive Gaussian noise and assume that devices are subject to individual transmit power constraints.
We propose a scheme where devices pick their repetition rates according to a prescribed distribution.
The devices then adapt their powers or select their rates as functions of the number of times they repeat their messages within a frame.
More specifically, devices with higher degrees transmit at lower power levels, a scheme which we call power adaptation (PA)-IRSA.
Alternatively, they can send information at higher rates under rate selection (RS)-IRSA.
In addition, the receiver uses MRC to recover sent messages rather than single slot decoding.
We demonstrate, through extensive numerical simulations, that the proposed schemes yield significant performance improvements, in terms of rate and power consumption, compared to existing schemes, especially in the low SNR regime.

\section{System Model}

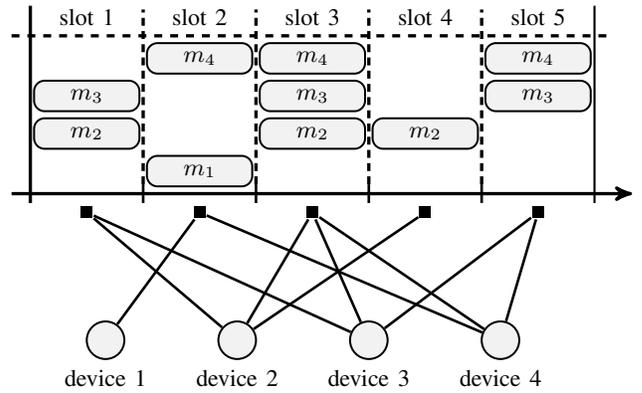
\begin{figure}[tb]
\centering{
\begin{tikzpicture}[
  line width=1pt, font=\small, >=stealth', draw=black,
  device/.style={circle, inner sep = 0pt, minimum height = 5mm, minimum width=5mm, draw=black, fill=gray!10, line width=0.75pt},
  slot/.style={rectangle, inner sep=0pt, minimum size=1.5mm, fill=black},
  packet/.style={rectangle, minimum height=4mm, minimum width=14mm, draw=black, fill=gray!10, rounded corners, line width=0.75pt}
]
\def\lw{1pt}

\draw [->,line width=1.25pt] (0,-0.05) -- (8.25,-0.05);
\draw[line width=1.25pt] (0.25,-0.175) -- (0.25,2.45);
\foreach \x in {1,...,4} {
  \draw (1.5*\x+0.25,-0.175) -- (1.5*\x+0.25,0.075);
  \draw[densely dashed,line width=1.25pt] (1.5*\x+0.25,0.075) -- (1.5*\x+0.25,2.40);
}
\draw[thick] (7.75,-0.175) -- (7.75,2.45);
\draw[dashed,line width=1.25pt] (0,2.05) -- (8,2.05);

\foreach \x in {1,...,5} {
  \node[slot] (s\x) at (1.5*\x-0.5,-0.3) {};
  \node (t\x) at (1.5*\x-0.5,2.3) {slot~\x};
}

\node[device] (d2) at (3,-2) [label=below:device~2]{};
\node[packet] (p24) at (5.5,0.75) {$m_2$};
\draw[line width=\lw] (d2) -- (s4);

\node[device] (d3) at (4.75,-2) [label=below:device~3]{};
\node[packet] (p31) at (1.0,1.25) {$m_3$};
\draw[line width=\lw] (d3) -- (s1);

\node[device] (d4) at (6.5,-2) [label=below:device~4]{};
\node[packet] (p43) at (4.0,1.75) {$m_4$};
\node[packet] (p45) at (7.0,1.75) {$m_4$};
\draw[line width=\lw] (d4) -- (s3);
\draw[line width=\lw] (d4) -- (s5);

\node[device] (d1) at (1.25,-2) [label=below:device~1]{};
\node[packet] (p12) at (2.5,0.25) {$m_1$};
\draw[line width=\lw] (d1) -- (s2);

\node[packet] (p21) at (1.0,0.75) {$m_2$};
\node[packet] (p23) at (4.0,0.75) {$m_2$};
\draw[line width=\lw] (d2) -- (s1);
\draw[line width=\lw] (d2) -- (s3);

\node[packet] (p33) at (4.0,1.25) {$m_3$};
\node[packet] (p35) at (7.0,1.25) {$m_3$};
\draw[line width=\lw] (d3) -- (s3);
\draw[line width=\lw] (d3) -- (s5);

\node[packet] (p42) at (2.5,1.75) {$m_4$};
\draw[line width=\lw] (d4) -- (s2);

\end{tikzpicture}}
\caption{This picture offers a notional diagram for the envision scheme.
In this instance, $K = 4$ devices are transmitting multiple replicas of their messages within a frame.
The messages associated with device~$i$ are labeled $m_i$.
The frame is partitioned into $M = 5$ slots.
Each device randomly selects the subset of slots during which they transmit their message.}
\label{figure:NotionalD}
\end{figure}

In this section, we introduce the abstract framework for the multiple access system we wish to study.
Let $K$ be the number of active devices, each of which is trying to transmit one short message to the receiver.
This communication task is enabled through a slotted structure, with every MAC frame consisting of $M$ slots.
Individual slots are composed of $\mathfrak{L}$ channel uses.
The focus is exclusively on the uplink scenario where information flows from the wireless devices to a central receiver.
%
We assume that the access point is aware of $K$, the total number of active devices, and it shares this information with each of them.
Based on the system parameters, every device randomly chooses a repetition pattern according to a prescribed distribution.
Specifically, device~$k$ independently draws repetition count $D_{k}$ using distribution $f_{D}$.
This device subsequently chooses $D_{k}$ slots uniformly within the frame, and then transmits its packet in all of the selected slots.
We note that this transmission patterns is established using common randomness between a user and the access point.
As a consequence, all patterns are known at the receiver.
A notional diagram for the proposed access scheme appears in Fig.~\ref{figure:NotionalD}.

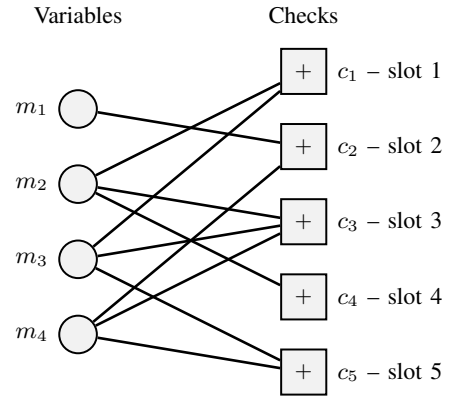
\begin{figure}[tb]
\centering{
\begin{tikzpicture}
  [
  font=\small, line width=1pt, draw=black,
  bitnode/.style={circle, inner sep = 0pt, minimum height = 5mm, minimum width=5mm, draw=black, fill=gray!10, line width=0.75pt},
  checknode/.style={rectangle, inner sep = 0pt, minimum size = 6mm, draw=black, fill=gray!10, line width=0.75pt},
  ]

\foreach \v in {1,2,3,4} {
  \node[bitnode] (v\v) at (0,0.5-\v) [label=left:{$m_\v$}]{};
}
  
\foreach \c in {1,2,3,4,5} {
  \node[checknode] (c\c) at (3,1-\c) [label=right:{$c_\c$ -- slot~\c}]{\footnotesize{$+$}};
}

\node (bit) at (0,0.75) {Variables};
\node (check) at (3,0.75) {Checks};

\draw (v1) -- (c2);
\draw (v2) -- (c1);
\draw (v2) -- (c3);
\draw (v2) -- (c4);
\draw (v3) -- (c1);
\draw (v3) -- (c3);
\draw (v3) -- (c5);
\draw (v4) -- (c2);
\draw (v4) -- (c3);
\draw (v4) -- (c5);
\end{tikzpicture}}
\caption{The envisioned random access scheme naturally admits a bipartite representation that can be put in correspondence with a Tanner graph.
In this analogy, the messages play the role of variable nodes, and the time slots act as check nodes.
Edges indicate the time slots during which a particular device transmits a copy of its message.
This relation provides a conceptual bridge between standard tools in iterative decoding and the multiple access problem at hand.
}
\label{fig:tangraph}
\end{figure}

The aggregate access schedule among the active devices can be represented as a bipartite graph, with messages on one side and slots on the other, as depicted in Fig.~\ref{fig:tangraph}.
When the receiver employs successive interference cancellation, this bipartite representation admits the application of powerful graphical tools from coding theory~\cite{liva2011graph}.
To leverage pertinent concepts from iterative decoding, we must first establish a suitable notation.
Following established literature, let $L(x)=\sum_{i} L_{i}x^i$, where $L_{i} = f_{D}(i)$, be the left degree distribution polynomial from the node perspective.
Likewise, $R(x)=\sum_{i} R_{i}x^i$ is the right degree distribution polynomial from the node perspective.
We emphasize that, once the variable node degree distribution is specified, the check node degree distribution is implicitly determined through the slot selection process.
The average variable node and check node degrees are given by
\begin{xalignat*}{2}
l_{\avg} &= \sum_i i L_{i} = L'(1)
& r_{\avg} &= \frac{K}{M} l_{\avg} ,
\end{xalignat*}
where function $L'(1)$ represents the derivative of polynomial $L(x)$ evaluated at one.

Herein, we are particularly interested in decoding strategies whereby the receiver uses maximal ratio combining, rather than single-slot decoding, to recover the sent messages sequentially.
Depending on the nature of the interference and background noise, such strategies can enhance performance drastically.
Additionally, we let devices adapt their power levels or select their code rates as functions of the degrees of their variable nodes, which correspond to the number of times a particular message is repeated within a frame.
In our envisioned implementation, users with higher degrees may transmit at lower power levels (PA-IRSA), or they can send information at higher rates (RS-IRSA).
A design challenge then is to find suitable left degree distributions for the nodes under various transmit power constraints.
We focus on low SNR scenarios, an operating regime where the impacts of the noise variance $N_{0}$ cannot be neglected.
This regime is common in sensor networks and machine-driven wireless communications.

\section{Rate Selection}

To establish a performance benchmark, we consider the situation where all the devices employ a same power level.
The transmit energy per channel use, $E_{s}$, is then uniform across slots and messages.
Moreover, the total expected energy expended by all devices within a frame is $E_{\tot} = K l_{\avg} \mathfrak{L} E_{s}$.
Suppose, in addition, that the decoding process relies on standard SIC via message-passing.
That is, MRC is not applied at the receiver at this point; only one replica of a message is used to recover the sent data.
Incidentally, this is the same procedure as the one proposed in IRSA~\cite{liva2011graph}.
Under this decoding strategy and for a message with degree $l_{i}$, the amount of energy transmitted by a device that is not employed during the decoding process is $\mathfrak{L}(l_{i}-1)E_{s}$.
This significantly reduces power efficiency, especially for degree distributions with high~$l_{\avg}$.
Under the standard information theoretic Gaussian approximation, this elementary scheme yields the following maximum available rate per user
\begin{equation*}
R_{\IRSA} = \frac{\mathfrak{L}}{2} \log \left( 1+\frac{E_{s}}{N_{0}} \right).
\end{equation*}
The sum of rates across all active devices per frame is then limited by
\begin{equation*}
\begin{split}
S_{\IRSA} &= M T_{\IRSA} R_{\IRSA} \\
&= M T_{\IRSA} \frac{\mathfrak{L}}{2} \log \left( 1+\frac{E_{s}}{N_{0}} \right) ,
\end{split}
\end{equation*}
where $T_{\IRSA}$ is the normalized throughput, i.e., the number of decoded messages within a frame divided by the total number of slots.
Equivalently, one can think of $M T_{\IRSA}$ as the total number of decoded packets via iterative message passing under standard SIC.

It is well known in the communication literature that MRC can improve decoding performance, especially in the low SNR regime \cite{brennan1959divers}.
This algorithmic opportunity therefore constitutes a promising means to enhance decoding performance for random multiple access schemes.
To illustrate these potential benefits, consider a particular realization of the frame.
We label the set of slots connected to user~$i$ by $U_i$.
Then, when we apply MRC at the receiver, the total SINR experienced by user~$i$ at iteration~$q$ becomes
\begin{equation*}
\sum_{j \in U_{i}} \frac{E_{s}}{(r_{j,q}-1)E_{s} + N_{0}} ,
\end{equation*}
where $r_{j,q}$ is the degree of slot~$j$ at decoding iteration~$q$ \cite{jakes1974microwave}.
We emphasize that the effective degree of a slot can change at every iteration of the decoding process because interference is being peeled off from some slots every time a message is recovered.
Then, under the Gaussian approximation, the maximum rate available to user~$i$ becomes
\begin{equation*}
R_{\max}^{i,q} = \frac{\mathfrak{L}}{2} \log \left( \sum_{j \in U_{i}} \frac{E_{s}}{(r_{j,q}-1)E_{s}+N_{0}} \right) .
\end{equation*}

While this argument is simple for a fixed realization, it becomes more complicated for a general system.
In the latter case, parameters $l_{i}$ and $r_{j,q}$ can be viewed as random variables, and they are determined by the construction of the transmission graph.
Performance improvements can accordingly be assessed in terms of expected rate, $\bar{R}_{\max} = \mathbb{E} \left[ R_{\max}^{i,q} \right]$.
Yet, the rate selection task by every device is more ambiguous.
In the MRC case, the maximum rate for a device depends on the realization of the bipartite graph.
That is, for decodability, we want every device to choose a rate slightly below $R_{\max}^{i,q}$, but this latter quantity depends on the actions of other users and the decoding order adopted by the receiver.
Unfortunately, for uncoordinated random access, wireless devices do not have access to the full transmission schedule and/or decoding order.
As such, they cannot select a maximum rate with certainty.

To circumvent this difficulty, we adopt a practical strategy and analyze its performance.
We assume that devices select their rate according to an estimated interference level, with the following formula
\begin{equation*}
R_{\RS}^{i} = \frac{\mathfrak{L}}{2} \log \left( 1 + \frac{E_{s}}{N_{0}} + \frac{\alpha (l_{i}-1) E_{s}}{(\beta r_{\avg} - 1) E_{s} + N_{0}} \right) .
\end{equation*}
Parameters $\alpha$ and $\beta$ are coefficients that depend on $M$, $K$, $E_{s}$, and the left degree distribution.
Naively, user~$i$ attempts to estimate $R_{\max}^{i,q}$ by assuming that one of the connected slots has degree one and all other slots have degree equal to $\beta r_{\avg}$.
The tuning variable $\alpha$ serves as a scaling parameter; it is chosen such that the selected rate remains close to $R_{\max}^{i,q}$.
Finding suitable values for $\alpha$ and $\beta$ is accomplished through numerical simulations.
From this preliminary analysis, we gather that wireless devices that transmit more frequently should pick higher information rates.
This is intuitively pleasing since devices that are expending more transmit energy within the context of a frame get larger rates.

\subsection{Decoding Process and Analysis}

We assume that the receiver has complete knowledge of the bipartite graph at the onset of the decoding process.
As such, it can compute $R_{\RS}^{i}$ for any device and predict whether a message can be successfully recovered at iteration~$q$.
Even when packet recovery is enhanced by MRC, the graphical portion of the decoding process follows a standard iterative message-passing algorithm.
That is, when there is a degree-one slot, the receiver attempts to decode the corresponding packet while leveraging its copies in other slots through MRC.
While the receiver decodes message~$i$, the effective SINR is determined by the bipartite graph resulting from peeling at this stage of the iterative process.
The decoding of message~$i$ succeeds and all replicas of this message are removed from the graph through interference cancellation whenever $R_{\RS}^{i}$ is less than the effective SINR at this particular stage.
This sequential process continues until there are no decodable degree-one slot left.
At this point, the receiver attempts to decode the remaining messages, although they may not be connected to degree-one slots.
Here again, MRC is applied and, as such, decoding succeeds whenever $R_{\RS}^{i}$ is below the maximum achievable rate under the residual SINR.
This cycle continues until no additional messages can be recovered through MRC.

\subsection{Efficiency Evaluation}

To compare different degree distributions, we explore two criteria.
First, adopting the approach put forth in \cite{liva2011graph}, we assess the efficiency of MAC schemes by normalizing their total rate over the sum capacity of the Gaussian MAC channel, $C_{\rf}$.
To make this notion clear, we go through an illustrative example.
Suppose that all the devices send packets in all the slots with constant energy $\tilde{E}_{s}$.
Then, the total energy transmitted by the users is $E_{\rf} = KM\mathfrak{L}\tilde{E}_{s}$ and the reference capacity per frame can be written as
\begin{equation*}
C_{\rf} = \frac{\mathfrak{L}M}{2}\log \left( 1+\frac{K\tilde{E}_{s}}{N_{0}} \right) .
\end{equation*}
We define the efficiency $\eta$ of a random access scheme as the ratio of the sum rate per frame over the reference capacity.
The fact that the total energy expended in the random MAC and reference schemes are equal, $E_{\tot} = E_{\rf}$, ensures a meaningful assessment of efficiency.
For our example, it is straightforward to show this balance is reached with
\begin{equation}\label{Enconst}
\tilde{E}_{s} = \frac{l_{\avg} E_{s}}{M} .
\end{equation}
One can then write the efficiency as
\begin{equation}\label{Eff}
\eta = \frac{S}{C_{\rf}}
= \frac{S}{\frac{\mathfrak{L}M}{2}\log \left( 1+\frac{K l_{\avg} {E}_{s}}{M N_{0}} \right)}
\end{equation}
where $S$ denotes the expected sum rate for all the devices within a frame.
Using \eqref{Enconst} and \eqref{Eff}, the efficiency associated with IRSA can be expressed as
\begin{equation*}\label{Effmp}
\begin{split}
\eta_{\IRSA}
= \frac{S_{\IRSA}}{C_{\rf}}& = T_{\IRSA}\frac{\log \left( 1+\frac{E_{s}}{N_{0}} \right)}{\log \left( 1+\frac{K l_{\avg} {E}_{s}}{M N_{0}} \right)} .
\end{split}
\end{equation*}

We can repeat these steps for a receiver that takes advantage of MRC.
For a prescribed degree distribution, let $\overline{R}_{\RS}$ be the expected value of the rate selected by individual users, i.e.,
\begin{equation*}
\begin{split}
\overline{R}_{\RS} &= \mathbb{E}[R_{\RS}^{i}] \\
&= \frac{\mathfrak{L}}{2} \mathbb{E} \left[ \log \left( 1 + \frac{E_{s}}{N_{0}} + \frac{\alpha (l_{i}-1) E_{s}}{(\beta r_{\avg} - 1) E_{s} + N_{0}} \right) \right] .
\end{split}
\end{equation*}
We note that we can derive a convenient upper bound for $\overline{R}_{\RS}$ using Jensen's inequality,
\begin{equation*}
\overline{R}_{\RS}
\leq \frac{\mathfrak{L}}{2}\log \left(1+\frac{E_{s}}{N_{0}}+\frac{\alpha (l_{\avg}-1) E_{s}}{(\beta r_{\avg}-1) E_{s} + N_{0}} \right) .
\end{equation*}
It is also worth mentioning that, in the low SNR regime, these two expressions are found empirically to be nearly indistinguishable.
With this in mind, the expected sum rate for all the users within a frame is given by
\begin{equation*}
\overline{S}_{\RS}
= \frac{\mathfrak{L}}{2} \mathbb{E} \left[ \sum_{i \in \mathcal{D}}
\log \left( 1 + \frac{E_{s}}{N_{0}} + \frac{\alpha (l_{i}-1) E_{s}}{(\beta r_{\avg} - 1) E_{s} + N_{0}} \right) \right]
\end{equation*}
where $\mathcal{D}$ is the set of decoded users.
Using the aforementioned approximation, we get
\begin{equation*}
\overline{S}_{\RS}
\approx \frac{MT_{\RS}\mathfrak{L}}{2}\log \left(1+\frac{E_{s}}{N_{0}}+\frac{\alpha (l_{\avg}-1) E_{s}}{(\beta r_{\avg}-1) E_{s} + N_{0}} \right)
\end{equation*}
where $T_{\RS} = |\mathcal{D}|/M$ is the expected throughput of the rate selection scheme.
Altogether, the efficiency of the MRC scheme with rate adaptation is adequately captured by
\begin{equation*}
\begin{split}
\eta_{\RS} &= \frac{\overline{S}_{\RS}}{C_{\rf}} \\
&\approx T_{\RS} \frac{\log \left( 1+\frac{E_{s}}{N_{0}}+\frac{\alpha (l_{\avg}-1) E_{s}}{(\beta r_{\avg}-1)E_{s}+N_{0}} \right)}{\log \left( 1 + \frac{K l_{\avg} E_{s}}{M N_{0}} \right)}.
\end{split}
\end{equation*}
Conceptually, rate selection is a means to take advantage of the benefits associated with MRC.
Ideally, we would like every device to select a rate matched to its effective SINR at the corresponding decoding step.
However and unfortunately, this is impractical in the uncoordinated framework.
Instead, devices choose rates close to $R_{\max}^{i}$ as a proxy for good performance.
To see how well $R_{\RS}^{i}$ performs compared to ideal rates, we introduce a maximum average efficiency,
\begin{equation*}
\begin{split}
\overline{S}_{\max} &= \mathbb{E} \left[ \sum_{i \in \mathcal{D}} R_{\max}^{i} \right]
\end{split}
\end{equation*}
%
%
%
%
%
Likewise, we can introduce the notion of maximum efficiency
\begin{align*}
\eta_{\RS,\max} = \frac{\overline{S}_{\max}}{C_{\rf}}.
\end{align*}
This quantity is pertinent in that it showcases how close the proposed framework can get to coordinated performance.
Furthermore, by comparing $\eta_{\RS}$ and $\eta_{\RS,\max}$, we can assess the suitability of our SINR prediction strategy.

Our second performance criterion is related to spectral efficiency.
This popular quantity indicates how much information is received per slot, on average.
For the message-passing decoder without SIC, the average spectral efficiency is proportional to
\begin{equation*}
\gamma_{\IRSA} = T_{\IRSA} R_{\IRSA}
\end{equation*}
Likewise, for MRC with rate adaptation, the average spectral efficiency can be written as
\begin{equation*}
\begin{split}
\gamma_{\RS}
&= \frac{\mathfrak{L}}{2 M} \mathbb{E} \left[ \sum_{i \in \mathcal{D}}
\log \left( 1 + \frac{E_{s}}{N_{0}} + \frac{\alpha (l_{i}-1) E_{s}}{(\beta r_{\avg} - 1) E_{s} + N_{0}} \right) \right].
\end{split}
\end{equation*}
%
A good benchmark for the maximum spectral efficiency rate selection with MRC is $\gamma_{\RS,\max} = \overline{S}_{\max}/M$.

Having discussed the potential benefits of rate adaptation, along with pertinent performance criteria, we turn to the second class of uncoordinated schemes we wish to study.
Specifically, in the next section, we explore the potential gains associated with varying transmit powers in the context of random access with MRC and SIC at the receiver.

\section{Power Adaptation}

For simplicity and fairness, assume that all the active devices adopt a common code rate to transmit data.
Then, owing to the random graph generation and the iterative decoding process, it is foreseeable that not all active devices need to employ a same transmit power.
Thus, we explore the benefits derived from power adaptation.

Let $\hat{R}$ denote the nominal rate shared by all active devices.
In the absence of interference and under the Gaussian approximation, the energy needed to sustain this rate, which we denote by $\hat{E}_{s}$, is implicitly given by
\begin{equation*}
\hat{R} = \frac{\mathfrak{L}}{2} \log \left( 1 + \frac{\hat{E}_{s}}{N_{0}} \right) .
\end{equation*}
In this section, we aim to design a power adaptation scheme such that the effective SINR of every user, after applying MRC and SIC at the receiver, is close to $\hat{E}_{s}/N_{0}$.

We note that this task is slightly more involved than rate selection.
While successful decoding for rate adaptation is dictated by the transmission graph, this information is not sufficient to accurately determine decodability under power adaptation.
The empirical distribution of the transmit powers associated with every slot is key in computing effective SINR.
Thus, we seek to estimate the right degree and the level of interference within each slot.

To gain insight, assume that the transmission schedule produces a regular bipartite graph.
That is, all the messages are repeated exactly $l_{\avg}$ times and every slot has degree $r_{\avg}$.
Also, suppose that the energy transmitted per channel use is $\bar{E}_{s}$ for all the active devices.
Then, the minimum possible SINR experienced during the iterative decoding process is
\begin{equation*}
\frac{l_{\avg} \bar{E}_{s}}{(r_{\avg}-1) \bar{E}_{s}+N_{0}} .
\end{equation*}
To find the value of $\bar{E}_{s}$ that would lead to a performance comparable to our idealized system, we equate the expression above to $\hat{E}_{s}/N_{0}$,
\begin{equation*}
\frac{\hat{E}_{s}}{N_{0}}
= \frac{l_{\avg} \bar{E}_{s}}{(r_{\avg}-1) \bar{E}_{s}+N_{0}} .
\end{equation*}
Solving for $\bar{E}_{s}$, we obtain
\begin{equation} \label{equation:BarES}
\bar{E}_{s}
= \frac{\hat{E}_{s}}{(1-r_{\avg})\frac{\hat{E}_{s}}{N_{0}}+l_{\avg}}.
\end{equation}
While unsophisticated, the argument above offers a starting point for power adaptation.

Consider a more general transmission schedule, with a possible irregular bipartite graph.
For the sake of argument, suppose that every device, other than device~$i$, transmits at nominal power $\bar{E}_{s}$ as defined in \eqref{equation:BarES}.
We wish to find conditions on $\check{E}_{s}^{i}$, the power used by device~$i$, such that the resulting SINR is equal to that of the idealized system.
Using the effective SINR associated with MRC, and by assuming that every slot contains $r_{\avg}$ messages, we get
\begin{equation} \label{esperchuse}
\frac{\hat{E}_{s}}{N_{0}}
= \frac{l_{i}\check{E}_{s}^{i}}{(r_{\avg}-1)\bar{E}_{s}+N_{0}}.
\end{equation}
Isolating the parameters associated with device~$i$, we obtain
\begin{equation*}
l_{i}\check{E}_{s}^{i}
= \frac{\hat{E}_{s}}{N_{0}} \left( (r_{\avg}-1)\bar{E}_{s}+N_{0} \right) .
\end{equation*}
This equation does not provide an optional power level for device~$i$.
Nevertheless, it offers a conceptual blueprint for a pragmatic solution.
In particular, in our candidate approach, we assume that device~$i$ uses a transmit power $E_{s}^{i}$ that is proportional to the quantity above;
\begin{equation*}
E_{s}^{i} = \mu \check{E}_{s}^{i} ,
\end{equation*}
where $\mu$ is a tuning parameters that depends on $M$, $K$, $\hat{R}$ and the left degree distribution.
The lowest value for $\mu$ that support decoding messages at a code rate equal to $\hat{R}$, is found through numerical computations.

Combining the expressions above, we gather that the total energy transmitted by device~$i$ is equal to
\begin{equation} \label{totesperuser}
\begin{split}
l_{i} E_{s}^{i} \mathfrak{L}
&= \mu \mathfrak{L} \frac{\hat{E}_{s}}{N_{0}} \left( (r_{\avg}-1)\bar{E}_{s}+N_{0} \right) \\
&= \mu \mathfrak{L} \hat{E}_{s}
\left( 1 + \frac{(r_{\avg}-1) \hat{E}_{s}}{(r_{\avg}-1) \hat{E}_{s}
+ N_{0} l_{\avg}} \right) .
\end{split}
\end{equation}
It is interesting to note that, irrespective of left degrees, the total energy used by the devices is the same over a frame.

\subsection{Decoding and Efficiency Evaluation}

As before, the receiver has complete information about the realization of the bipartite transmission graph.
Consequently, it can compute the transmit power selected by every device.
With this knowledge, the iterative decoding algorithm proceeds in a manner akin to the standard process.
The receiver initially attempts to decode devices that are connected to degree-one slots.
When the effective SINR of a message is greater than $\hat{E}_{s}/N_{0}$, decoding succeeds.
The corresponding data is recovered and the codeword can be peeled.
Once this phase terminates, the receiver tries to decode the remaining messages based on the same SINR criteria, even when devices are not connected to degree-one slots.

To evaluate the performance associated with different degree distributions, we explore the average energy per device normalized by $N_{0}$.
More precisely, we compare the average normalized energy of the proposed power adaptation scheme
\begin{equation*}
\begin{split}
\Gamma_{\PA}
= \mu \frac{\hat{E}_{s}}{N_{0}}
\left( N_{0} + \frac{(r_{\avg}-1) \hat{E}_{s} N_{0}}{(r_{\avg}-1) \hat{E}_{s}
+ N_{0} l_{\avg}} \right)
\end{split}
\end{equation*}
to the normalized energy associated with IRSA and the minimum level derived from the idealized central scheme,
\begin{xalignat*}{2}
\Gamma_{\IRSA} &= l_{\avg} \frac{\hat{E}_{s}}{N_{0}} &
\Gamma_{\min} &= \frac{\hat{E}_{s}}{N_{0}} ,
\end{xalignat*}
respectively.
Coefficient $\mathfrak{L}$ is omitted from these definitions because it is common to all of them and, hence, does not influence the relative character of these values.

Finally, we need to establish a means to fairly compare the performance of the proposed rate selection and power adaptation schemes.
For a fixed ratio $K/M$, this is accomplished by first identifying the maximum rate at which rate selection can operate.
We then adopt this rate as a target value for power adaptation, and see if it can be attained using less energy.

\begin{figure}[t]
%
%
%
\definecolor{mycolor1}{rgb}{1.00000,0.00000,1.00000}%
\definecolor{mycolor2}{rgb}{0.00000,1.00000,1.00000}%
\begin{tikzpicture}

\begin{axis}[%
width=0.85\columnwidth,
height=0.7417\columnwidth,
scale only axis,
every outer x axis line/.append style={white!15!black},
every x tick label/.append style={font=\color{white!15!black}},
xmin=0,
xmax=1.55,
xlabel={G},
xmajorgrids,
every outer y axis line/.append style={white!15!black},
every y tick label/.append style={font=\color{white!15!black}},
ymin=0,
ymax=1.75,
ylabel={Normalized Throughput (T)},
ymajorgrids,
axis x line*=bottom,
axis y line*=left,
legend entries={$L^{(1)}(x)$ - RS-IRSA, $L^{(3)}(x)$ - RS-IRSA, $L^{(2)}(x)$ - RS-IRSA, $L^{(1)}(x)$ - IRSA, $L^{(3)}(x)$ - IRSA, $L^{(2)}(x)$ - IRSA},
legend style={at={(0.03,0.97)},anchor=north west,draw=white!15!black,fill=white,legend cell align=left,nodes={scale=0.7, transform shape}}
]
\addplot [color=blue,line width=1.0pt,mark size=3.0pt,only marks,mark=triangle,mark options={solid}]
  table[row sep=crcr]{0.05	0.049984503\\
0.1	0.099924692\\
0.15	0.149925\\
0.2	0.199734843\\
0.25	0.249584513\\
0.3	0.299684\\
0.35	0.349261072\\
0.4	0.398992011\\
0.45	0.449061469\\
0.5	0.497058236\\
0.55	0.54782967\\
0.6	0.595934\\
0.65	0.642997835\\
0.7	0.687965035\\
0.75	0.740970075\\
0.8	0.787571809\\
0.85	0.84129745\\
0.9	0.885727545\\
0.95	0.933702532\\
1	0.982245847\\
1.05	1.028653846\\
1.1	1.079432234\\
1.15	1.12856705\\
1.2	1.177692\\
1.25	1.223149378\\
1.3	1.272307359\\
1.35	1.32129148\\
1.4	1.367139535\\
1.45	1.421043478\\
1.5	1.468039801\\
};
\addplot [color=mycolor1,line width=1.0pt,mark size=3pt,only marks,mark=diamond,mark options={solid}]
  table[row sep=crcr]{0.05	0.049936344\\
0.1	0.099508164\\
0.15	0.1484515\\
0.2	0.199633578\\
0.25	0.249198168\\
0.3	0.299909\\
0.35	0.349486014\\
0.4	0.399154461\\
0.45	0.449313343\\
0.5	0.499078203\\
0.55	0.549357143\\
0.6	0.599894\\
0.65	0.649114719\\
0.7	0.6992331\\
0.75	0.748049875\\
0.8	0.797710106\\
0.85	0.847093484\\
0.9	0.893107784\\
0.95	0.943253165\\
1	0.988830565\\
1.05	1.041762238\\
1.1	1.087787546\\
1.15	1.138279693\\
1.2	1.190856\\
1.25	1.232514523\\
1.3	1.287883117\\
1.35	1.332076233\\
1.4	1.383930233\\
1.45	1.43552657\\
1.5	1.480308458\\
};
\addplot [color=red,line width=1.0pt,mark size=3pt,only marks,mark=o,mark options={solid}]
  table[row sep=crcr]{0.05	0.049914514\\
0.1	0.099482839\\
0.15	0.1482555\\
0.2	0.195892738\\
0.25	0.249079101\\
0.3	0.29865\\
0.35	0.347623543\\
0.4	0.395850866\\
0.45	0.449121439\\
0.5	0.498264559\\
0.55	0.547694139\\
0.6	0.597736\\
0.65	0.645989177\\
0.7	0.698773893\\
0.75	0.747022444\\
0.8	0.795452128\\
0.85	0.84403966\\
0.9	0.893188623\\
0.95	0.940531646\\
1	0.988305648\\
1.05	1.040625874\\
1.1	1.086468864\\
1.15	1.137042146\\
1.2	1.186504\\
1.25	1.229518672\\
1.3	1.283688312\\
1.35	1.328403587\\
1.4	1.380390698\\
1.45	1.43021256\\
1.5	1.477631841\\
};
\addplot [color=black,line width=1.0pt,mark size=3pt,only marks,mark=asterisk,mark options={solid}]
  table[row sep=crcr]{0.05	0.0499916680553\\
0.1	0.0999660113296\\
0.15	0.149991\\
0.2	0.199850099933\\
0.25	0.249752706078\\
0.3	0.299942\\
0.35	0.349548951049\\
0.4	0.399287616511\\
0.45	0.449452773613\\
0.5	0.497216306156\\
0.55	0.545018315018\\
0.6	0.592544\\
0.65	0.632569264069\\
0.7	0.671221445221\\
0.75	0.692992518703\\
0.8	0.681736702128\\
0.85	0.597\\
0.9	0.43074251497\\
0.95	0.170329113924\\
1	0.0659102990033\\
1.05	0.030951048951\\
1.1	0.0170769230769\\
1.15	0.0116283524904\\
1.2	0.008244\\
1.25	0.00636514522822\\
1.3	0.00437229437229\\
1.35	0.00305381165919\\
1.4	0.00246511627907\\
1.45	0.00184057971014\\
1.5	0.00129353233831\\
};
\addplot [color=green,line width=1.0pt,mark size=3pt,only marks,mark=+,mark options={solid}]
  table[row sep=crcr]{0.05	0.0499913347775\\
0.1	0.0999623458847\\
0.15	0.149994\\
0.2	0.199850099933\\
0.25	0.249755203997\\
0.3	0.299944\\
0.35	0.349534965035\\
0.4	0.399299600533\\
0.45	0.44952173913\\
0.5	0.498727121464\\
0.55	0.548833333333\\
0.6	0.598972\\
0.65	0.647924242424\\
0.7	0.697179487179\\
0.75	0.744713216958\\
0.8	0.787449468085\\
0.85	0.803841359773\\
0.9	0.646038922156\\
0.95	0.288259493671\\
1	0.144019933555\\
1.05	0.0912587412587\\
1.1	0.0673223443223\\
1.15	0.0508659003831\\
1.2	0.039284\\
1.25	0.0323858921162\\
1.3	0.0260649350649\\
1.35	0.0215022421525\\
1.4	0.0173441860465\\
1.45	0.0134927536232\\
1.5	0.0118457711443\\
};
\addplot [color=mycolor2,line width=1.0pt,mark size=3pt,only marks,mark=x,mark options={solid}]
  table[row sep=crcr]{0.05	0.049990668222\\
0.1	0.099965344885\\
0.15	0.1499965\\
0.2	0.199848767488\\
0.25	0.249751873439\\
0.3	0.299914\\
0.35	0.349546620047\\
0.4	0.399271637816\\
0.45	0.449503748126\\
0.5	0.498695507488\\
0.55	0.54871978022\\
0.6	0.598982\\
0.65	0.647649350649\\
0.7	0.696869463869\\
0.75	0.743815461347\\
0.8	0.788484042553\\
0.85	0.77533427762\\
0.9	0.522149700599\\
0.95	0.295091772152\\
1	0.208624584718\\
1.05	0.163664335664\\
1.1	0.129882783883\\
1.15	0.105233716475\\
1.2	0.088548\\
1.25	0.075755186722\\
1.3	0.0621212121212\\
1.35	0.0541614349776\\
1.4	0.0454046511628\\
1.45	0.0381014492754\\
1.5	0.0337064676617\\
};
\end{axis}
\end{tikzpicture}%
\caption{Throughput, $T$, of RS-IRSA and IRSA with ideal soliton ($L^{(1)}(x)$) with parameter $M$ (number of slots), modified soliton ($L^{(2)}(x)$) with parameter 10 and $L^{(3)}(x)$ as left degree distributions. $K=300, \frac{\tilde{E}_{s}}{N_{o}}=0.0009$.}
\label{fig:rtthrok3h}
\end{figure}
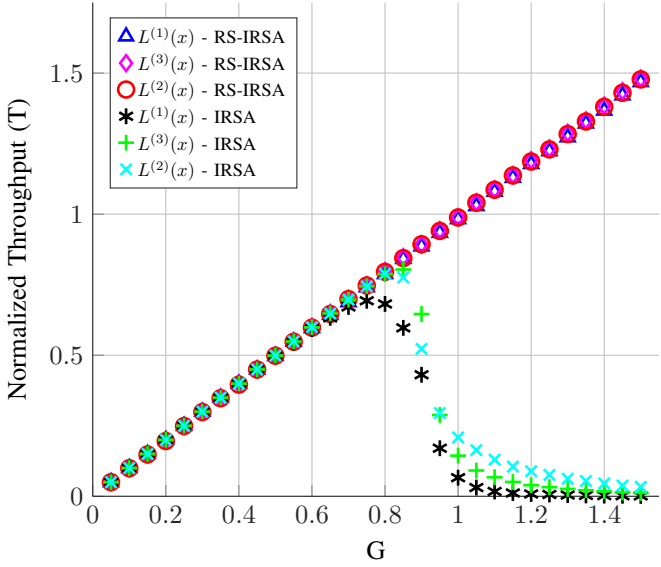

\section{Numeric Results}

Since there are no closed-form expressions for throughput as a function of $\alpha$ and $\beta$ for rate selection, nor as a function of $\mu$ for power adaptation, it is difficult to establish optimal left degree distributions for these algorithms, both for finite $K$ and asymptotically.
Instead, we seek to demonstrate the value of the proposed schemes over existing ones.
This is accomplished through numerical computations with three expository degree distributions.
Note that, in all of them, almost half of the variable nodes have degree equal to two.
This conforms to the intuition that, when properly accounting for radio resources, excessive repetitions are costly.
The candidate left degree distributions are as follows:
\begin{itemize}
\item The ideal soliton, which is deployed in \cite{madalaicc2015}, with parameter $Y$,
\begin{equation*}
L_{i}^{(1)}= \begin{cases} \frac{1}{Y}, & \ i=1 \\
\frac{1}{i(i-1)}, & \ 2\leq i\leq Y ; \end{cases}
\end{equation*}
\item Our proposed distribution, the modified soliton with parameter $Y$,
\begin{equation*}
L_{i}^{(2)}=\frac{1}{i(i-1)}+\frac{1}{Y(Y-1)}, \quad 2\leq i\leq Y ;
\end{equation*}
\item And, the numerically validated discrete distribution, which is used in \cite{liva2011graph},
\begin{equation*}
\begin{split}
&L^{(3)} = 0.4977x^{2} + 0.2207x^{3} + 0.0381x^{4} + 0.0756x^{5} \\
&+ 0.0398x^{6} + 0.0009x^{7} + 0.0088x^{8} + 0.0068x^{9} \\
&+ 0.0030x^{11} + 0.0429x^{14} + 0.0081x^{15} + 0.0576x^{16} .
\end{split}
\end{equation*}
\end{itemize}

\subsection{Rate Selection}

\begin{figure}[t]
%
%
%
\definecolor{mycolor1}{rgb}{1.00000,0.00000,1.00000}%
\definecolor{mycolor2}{rgb}{0.00000,1.00000,1.00000}%

\begin{tikzpicture}

\begin{axis}[%
width=0.85\columnwidth,
height=0.7417\columnwidth,
scale only axis,
every outer x axis line/.append style={white!15!black},
every x tick label/.append style={font=\color{white!15!black}},
xmin=0,
xmax=1.55,
xlabel={G},
xmajorgrids,
every outer y axis line/.append style={white!15!black},
every y tick label/.append style={font=\color{white!15!black}},
ymin=0,
ymax=1.3,
ylabel={$\text{Efficiency (}\eta\text{)}$},
ymajorgrids,
axis x line*=bottom,
axis y line*=left,
legend columns=3,
legend entries={$L^{(1)}(x)$ - $\eta_{RS}$, $L^{(3)}(x)$ - $\eta_{RS}$, $L^{(2)}(x)$-$\eta_{RS}$, $L^{(1)}(x)$-$\eta_{IRSA}$, $L^{(3)}(x)$ - $\eta_{IRSA}$, $L^{(2)}(x)$ - $\eta_{IRSA}$, $L^{(1)}(x)$ - $\eta_{RS,max}$, $L^{(3)}(x)$ - $\eta_{RS,max}$, $L^{(2)}(x)$ - $\eta_{RS,max}$},
legend style={at={(0.03,0.97)},anchor=north west,draw=white!15!black,fill=white,legend cell align=left,nodes={scale=0.6, transform shape}}
]
\addplot [color=red,line width=1.0pt,mark size=3.0pt,only marks,mark=triangle,mark options={solid}]
  table[row sep=crcr]{0.05	0.156806203174604\\
0.1	0.256184116455851\\
0.15	0.316637981378809\\
0.2	0.360520204108379\\
0.25	0.401230495350503\\
0.3	0.433660060602404\\
0.35	0.463285672245916\\
0.4	0.488490126380294\\
0.45	0.513414045928792\\
0.5	0.532768751976589\\
0.55	0.550987266355532\\
0.6	0.569473448880298\\
0.65	0.582487180127292\\
0.7	0.580271002092503\\
0.75	0.538331958751636\\
0.8	0.534012382303666\\
0.85	0.524492706782895\\
0.9	0.52805735644058\\
0.95	0.534807250181932\\
1	0.53986535018652\\
1.05	0.544830043873047\\
1.1	0.551503488330155\\
1.15	0.553395325150848\\
1.2	0.560128040615367\\
1.25	0.56561907529112\\
1.3	0.570073197612727\\
1.35	0.574312182394327\\
1.4	0.581007800151813\\
1.45	0.582185746853369\\
1.5	0.588049393360699\\
};

\addplot [color=black,line width=1.0pt,mark size=3pt,only marks,mark=diamond,mark options={solid}]
  table[row sep=crcr]{0.05	0.250130480838148\\
0.1	0.376111422397971\\
0.15	0.439518387668307\\
0.2	0.493188408499365\\
0.25	0.533681059940707\\
0.3	0.565198814594345\\
0.35	0.591323582352386\\
0.4	0.611363988399004\\
0.45	0.631986809892752\\
0.5	0.649505198975933\\
0.55	0.66632877932336\\
0.6	0.680404185287607\\
0.65	0.693410946047646\\
0.7	0.706148915845952\\
0.75	0.715970616387843\\
0.8	0.72467317526156\\
0.85	0.720492530579271\\
0.9	0.654978623792568\\
0.95	0.631597863501624\\
1	0.626238567564901\\
1.05	0.613495297898529\\
1.1	0.610066242540902\\
1.15	0.612690571264842\\
1.2	0.617410648741919\\
1.25	0.616213443251098\\
1.3	0.620109186174469\\
1.35	0.619778511763245\\
1.4	0.622486614964265\\
1.45	0.624173659483329\\
1.5	0.624140143103197\\
};

\addplot [color=blue,line width=1.0pt,mark size=3pt,only marks,mark=o,mark options={solid}]
  table[row sep=crcr]{0.05	0.282654967085139\\
0.1	0.414944926229741\\
0.15	0.480580759776655\\
0.2	0.519489893563085\\
0.25	0.578531545442874\\
0.3	0.607807432179725\\
0.35	0.630743589980286\\
0.4	0.649991068943281\\
0.45	0.677411570458422\\
0.5	0.694595215105803\\
0.55	0.707481874776871\\
0.6	0.720816278373705\\
0.65	0.730328318592442\\
0.7	0.748280273537357\\
0.75	0.757216551266989\\
0.8	0.765171567746029\\
0.85	0.75716052257404\\
0.9	0.695806271648357\\
0.95	0.669326448953457\\
1	0.664095603333195\\
1.05	0.654360560409532\\
1.1	0.649026442119587\\
1.15	0.65182041624911\\
1.2	0.652684592819545\\
1.25	0.652119785125997\\
1.3	0.655057330433298\\
1.35	0.654688526716862\\
1.4	0.658383187495903\\
1.45	0.658111072233549\\
1.5	0.661297723429804\\
};

\addplot [color=mycolor1,line width=1.0pt,mark size=3pt,only marks,mark=asterisk,mark options={solid}]
  table[row sep=crcr]{0.05	0.101007324257\\
0.1	0.121571426434\\
0.15	0.133375058422\\
0.2	0.141718380599\\
0.25	0.148249356507\\
0.3	0.15369092401\\
0.35	0.158304013067\\
0.4	0.162372459468\\
0.45	0.166038650887\\
0.5	0.168813950465\\
0.55	0.171195262143\\
0.6	0.17330931019\\
0.65	0.173537360952\\
0.7	0.173413103709\\
0.75	0.169517659047\\
0.8	0.158301948898\\
0.85	0.131729671139\\
0.9	0.0908905853577\\
0.95	0.03436952487\\
1	0.012788167036\\
1.05	0.0057629752026\\
1.1	0.00306282447308\\
1.15	0.00201157158863\\
1.2	0.00137764015453\\
1.25	0.00103283203288\\
1.3	0.000685768405098\\
1.35	0.000465642760892\\
1.4	0.000365052401264\\
1.45	0.000264430934353\\
1.5	0.000181525128796\\
};

\addplot [color=green,line width=1.0pt,mark size=3pt,only marks,mark=+,mark options={solid}]
  table[row sep=crcr]{0.05	0.171773866058\\
0.1	0.206078883934\\
0.15	0.221993869814\\
0.2	0.2311934462\\
0.25	0.237216106781\\
0.3	0.241489473708\\
0.35	0.244605184208\\
0.4	0.247029423124\\
0.45	0.248959919841\\
0.5	0.250453620605\\
0.55	0.251724559298\\
0.6	0.252702961908\\
0.65	0.253521364359\\
0.7	0.25413262518\\
0.75	0.254445291268\\
0.8	0.252900565765\\
0.85	0.242929495016\\
0.9	0.185082741734\\
0.95	0.0782737249945\\
1	0.0373069691003\\
1.05	0.0224956516634\\
1.1	0.015861741229\\
1.15	0.011471608733\\
1.2	0.00849567362052\\
1.25	0.00675791187129\\
1.3	0.0052185593476\\
1.35	0.00415934606244\\
1.4	0.00323730599816\\
1.45	0.00242670905076\\
1.5	0.00207001284552\\
};

\addplot [color=mycolor2,line width=1.0pt,mark size=3pt,only marks,mark=x,mark options={solid}]
  table[row sep=crcr]{0.05	0.197832065862\\
0.1	0.242958771897\\
0.15	0.264800138986\\
0.2	0.277722961105\\
0.25	0.28631507862\\
0.3	0.292451712765\\
0.35	0.297025216093\\
0.4	0.300542138822\\
0.45	0.303394270813\\
0.5	0.305608820915\\
0.55	0.307459187745\\
0.6	0.309017244792\\
0.65	0.310128709631\\
0.7	0.311088347612\\
0.75	0.311421422216\\
0.8	0.310479981148\\
0.85	0.287429389228\\
0.9	0.18357629977\\
0.95	0.0983733495472\\
1	0.0663691718728\\
1.05	0.0495630576422\\
1.1	0.0376054400872\\
1.15	0.0291728066979\\
1.2	0.0235448245565\\
1.25	0.0194398551039\\
1.3	0.0152988076979\\
1.35	0.0128894735085\\
1.4	0.0104283150895\\
1.45	0.0084337960324\\
1.5	0.00725016618885\\
};

\addplot [color=purple,line width=1.0pt,mark size=3pt,only marks,mark=star,mark options={solid}]
  table[row sep=crcr]{0.05	0.244541699177662\\
0.1	0.340168608135387\\
0.15	0.406840135820805\\
0.2	0.457579482610345\\
0.25	0.499322449526368\\
0.3	0.535141596118901\\
0.35	0.567102235905899\\
0.4	0.593546385664816\\
0.45	0.618996651607067\\
0.5	0.637825710059569\\
0.55	0.655552076933022\\
0.6	0.672568651438032\\
0.65	0.686543984054912\\
0.7	0.693701748948625\\
0.75	0.716688725456708\\
0.8	0.723758360895217\\
0.85	0.738010792585661\\
0.9	0.742513902871399\\
0.95	0.749604640497207\\
1	0.758033484083086\\
1.05	0.763498560787897\\
1.1	0.773634842789204\\
1.15	0.777379229300466\\
1.2	0.787420284208497\\
1.25	0.797099399433897\\
1.3	0.803247328757027\\
1.35	0.811822688350904\\
1.4	0.82162771046758\\
1.45	0.824041949697128\\
1.5	0.834531309374714\\
};

\addplot [color=gray,line width=1.0pt,mark size=3pt,only marks,mark=otimes,mark options={solid}]
  table[row sep=crcr]{0.05	0.357325221713009\\
0.1	0.491200309986215\\
0.15	0.567552996711183\\
0.2	0.614776358155501\\
0.25	0.675724942909312\\
0.3	0.707902866873621\\
0.35	0.731831688707049\\
0.4	0.750887687816443\\
0.45	0.778100926850789\\
0.5	0.794395552767649\\
0.55	0.805028405949584\\
0.6	0.816166319366655\\
0.65	0.823469623227397\\
0.7	0.839557378996599\\
0.75	0.846265490416368\\
0.8	0.851578450438513\\
0.85	0.853416359659688\\
0.9	0.861078563704715\\
0.95	0.859123344282291\\
1	0.863762826376206\\
1.05	0.871707211438477\\
1.1	0.869327969670056\\
1.15	0.87444001409908\\
1.2	0.876693821666953\\
1.25	0.878237748443371\\
1.3	0.882505235882074\\
1.35	0.883914189235495\\
1.4	0.889788854985067\\
1.45	0.890012182812385\\
1.5	0.896103592063878\\
};

\addplot [color=yellow,line width=1.0pt,mark size=3pt,only marks,mark=square,mark options={solid}]
  table[row sep=crcr]{0.05	0.336751823663289\\
0.1	0.460810914894032\\
0.15	0.532889187570155\\
0.2	0.596030670204993\\
0.25	0.636813019956337\\
0.3	0.671521973924856\\
0.35	0.699568011314278\\
0.4	0.719878913979953\\
0.45	0.74015963802114\\
0.5	0.756988881105967\\
0.55	0.772443361955313\\
0.6	0.784588018628747\\
0.65	0.795835763983891\\
0.7	0.806558805272246\\
0.75	0.814384856138161\\
0.8	0.820766560898617\\
0.85	0.826846876471394\\
0.9	0.830283169197174\\
0.95	0.831820117778296\\
1	0.836225174681077\\
1.05	0.840379414351764\\
1.1	0.840902705935873\\
1.15	0.845895215896412\\
1.2	0.853745152159591\\
1.25	0.854779084543193\\
1.3	0.860559793664325\\
1.35	0.862410090251902\\
1.4	0.867387043730681\\
1.45	0.870388041856634\\
1.5	0.872608945801979\\
};

\end{axis}
\end{tikzpicture}%
\caption{Efficiency, $\eta$, of RS-IRSA and IRSA with ideal soliton ($L^{(1)}(x)$) with parameter $M$ (number of slots), modified soliton ($L^{(2)}(x)$) with parameter 10 and $L^{(3)}(x)$ as left degree distributions. $K=300, \frac{\tilde{E}_{s}}{N_{o}}=0.0009$.}
\label{fig:effck3h}
\end{figure}
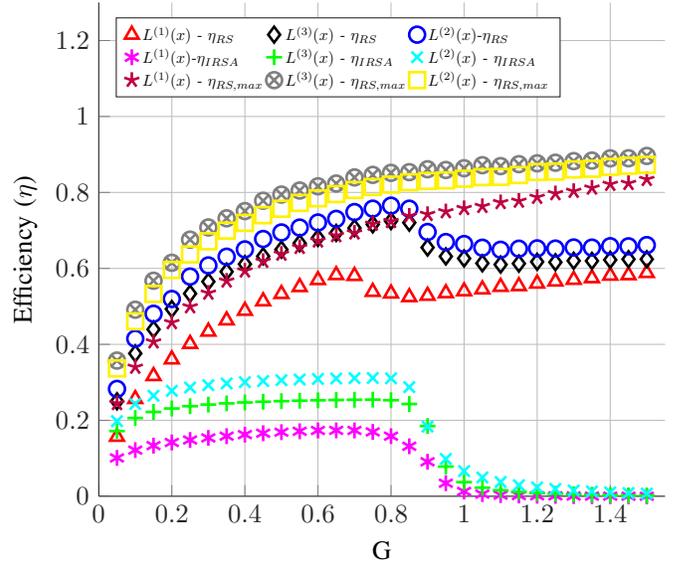

Because we are primarily interested in performance at low SNR, simulations are performed at
\begin{equation*}
\frac{\tilde{E}_{s}}{N_{0}}
= 0.0009 = -30.46 \text{ dB}.
\end{equation*}
We note that this puts $E_{s}/N_{0}$ in the range of -12.78~dB to 1.97~dB.
Figure~\ref{fig:effck3h} plots the efficiency $\eta$ of RS-IRSA and IRSA for different values of $M$, while $K$ remains constant at $300$.
Parameters $\alpha$ and $\beta$ are derived numerically through simulations as to ensure that throughput is almost equal to $G = K/M$ for the three expository distributions.
The results are averaged over 1000 independent trials for every operating point.
As mentioned above, one may anticipate that distributions with higher $l_{\avg}$ will show lower performance because of their misuse of radio resources.
There is a natural tension between a desire for a high left degree to boost performance through MRC, and the fact that higher slot counts also produce more interference.

For rate selection, the modified soliton shows the best performance.
This can be attributed to the fact that it features the lowest average left degree $l_{\avg}^{(2)} = 3.42$ among the three distributions.
On the other hand, the ideal soliton offers the lowest performance because of its high average degree, $l_{\avg}^{(1)} = 5.7$.
Our scheme is very close to the maximum possible rate for $G<0.8$, only losing $0.1$ in terms of efficiency.
However, when $G>0.8$, efficiency degrades because most slots have high right degrees and a user with a low left degree cannot support a high rate in the presence of many interferers.
To maintain a good estimation for $G>0.8$, we must decrease $\alpha$ or increase $\beta$, which leads to a reduction in the sum capacity of the frame.
We stress that, in the aforementioned setup, RS-IRSA can reach up to 80\% of the capacity associated with the best coordinated scheme, and it features a 50\% improvement in terms of capacity with respect to IRSA.
Altogether, the benefits of the proposed schemes are very significant.

\begin{figure}[t]
%
%
%
\definecolor{mycolor1}{rgb}{1.00000,0.00000,1.00000}%
\definecolor{mycolor2}{rgb}{0.00000,1.00000,1.00000}%
\begin{tikzpicture}

\begin{axis}[%
width=0.85\columnwidth,
height=0.7417\columnwidth,
scale only axis,
every outer x axis line/.append style={white!15!black},
every x tick label/.append style={font=\color{white!15!black}},
xmin=0,
xmax=1.55,
xlabel={G},
xmajorgrids,
every outer y axis line/.append style={white!15!black},
every y tick label/.append style={font=\color{white!15!black}},
ymin=0,
ymax=21,
ylabel={$\text{Spectral Efficiency (}\gamma\ \text{bit/slot)}$},
ymajorgrids,
axis x line*=bottom,
axis y line*=left,
legend columns=3,
legend entries={$L^{(1)}(x)$ - $\gamma_{RS}$, $L^{(3)}(x)$ - $\gamma_{RS}$, $L^{(2)}(x)$-$\gamma_{RS}$, $L^{(1)}(x)$-$\gamma_{IRSA}$, $L^{(3)}(x)$ - $\gamma_{IRSA}$, $L^{(2)}(x)$ - $\gamma_{IRSA}$, $L^{(1)}(x)$ - $\gamma_{RS,max}$, $L^{(3)}(x)$ - $\gamma_{RS,max}$, $L^{(2)}(x)$ - $\gamma_{RS,max}$},
legend style={at={(0.03,0.97)},anchor=north west,draw=white!15!black,fill=white,legend cell align=left,nodes={scale=0.6, transform shape}}
]
\addplot [color=red,line width=1.0pt,mark size=3.0pt,only marks,mark=triangle,mark options={solid}]
  table[row sep=crcr]{0.05	2.70356238958067\\
0.1	4.41697920141597\\
0.15	5.45928995901525\\
0.2	6.21588199269572\\
0.25	6.91778543610603\\
0.3	7.47691734326152\\
0.35	7.98770509331137\\
0.4	8.42226581106128\\
0.45	8.85198968680885\\
0.5	9.18569239868416\\
0.55	9.49980554136732\\
0.6	9.8185336750962\\
0.65	10.042908931475\\
0.7	10.0046988775466\\
0.75	9.28161001237135\\
0.8	9.20713436474712\\
0.85	9.04300158216937\\
0.9	9.10446122864129\\
0.95	9.22083901742409\\
1	9.30804786475385\\
1.05	9.39364625395893\\
1.1	9.50870594585523\\
1.15	9.54132391610654\\
1.2	9.65740551752734\\
1.25	9.75207878299118\\
1.3	9.82887418658876\\
1.35	9.90196033826984\\
1.4	10.0174023312193\\
1.45	10.0377118042239\\
1.5	10.1388094207538\\
};
\addplot [color=black,line width=1.0pt,mark size=3pt,only marks,mark=diamond,mark options={solid}]
  table[row sep=crcr]{0.05	4.31260589425286\\
0.1	6.48469682962621\\
0.15	7.57792324642636\\
0.2	8.50327090165061\\
0.25	9.20142188528879\\
0.3	9.74483288445342\\
0.35	10.1952611073627\\
0.4	10.5407862655959\\
0.45	10.8963530837303\\
0.5	11.1983950791827\\
0.55	11.4884575781818\\
0.6	11.731137630812\\
0.65	11.9553927200077\\
0.7	12.1750134676891\\
0.75	12.3443535747127\\
0.8	12.4943980826898\\
0.85	12.4223178126052\\
0.9	11.2927647114349\\
0.95	10.8896471048981\\
1	10.7972452019648\\
1.05	10.5775330710502\\
1.1	10.5184112704568\\
1.15	10.5636584401584\\
1.2	10.6450393006387\\
1.25	10.624397777852\\
1.3	10.6915659295681\\
1.35	10.6858646331686\\
1.4	10.732556195476\\
1.45	10.7616432497568\\
1.5	10.7610653721288\\
};
\addplot [color=blue,line width=1.0pt,mark size=3pt,only marks,mark=o,mark options={solid}]
  table[row sep=crcr]{0.05	4.87337434744547\\
0.1	7.15424175795015\\
0.15	8.28589705164221\\
0.2	8.95674594957339\\
0.25	9.97470817136382\\
0.3	10.479466165806\\
0.35	10.874918205825\\
0.4	11.2067721780813\\
0.45	11.6795406724248\\
0.5	11.9758112021057\\
0.55	12.197995577624\\
0.6	12.427899688637\\
0.65	12.5919008159966\\
0.7	12.9014181051574\\
0.75	13.0554922641585\\
0.8	13.1926480846989\\
0.85	13.0545262548899\\
0.9	11.9966915422683\\
0.95	11.5401416797592\\
1	11.4499544424694\\
1.05	11.2821084270142\\
1.1	11.1901406163697\\
1.15	11.2383127230566\\
1.2	11.2532123603394\\
1.25	11.2434742735329\\
1.3	11.2941217340214\\
1.35	11.2877630302202\\
1.4	11.3514642562193\\
1.45	11.3467725913071\\
1.5	11.4017150078974\\
};
\addplot [color=mycolor1,line width=1.0pt,mark size=3pt,only marks,mark=asterisk,mark options={solid}]
  table[row sep=crcr]{0.05	1.74151019046\\
0.1	2.09606461275\\
0.15	2.29957604663\\
0.2	2.44342680894\\
0.25	2.55603013925\\
0.3	2.64985051642\\
0.35	2.72938674473\\
0.4	2.79953255761\\
0.45	2.86274292144\\
0.5	2.91059304055\\
0.55	2.95165024689\\
0.6	2.98809944743\\
0.65	2.99203136749\\
0.7	2.98988899557\\
0.75	2.92272597919\\
0.8	2.72935115551\\
0.85	2.27120722545\\
0.9	1.56708319701\\
0.95	0.592579580172\\
1	0.22048622092\\
1.05	0.0993619038674\\
1.1	0.0528074579809\\
1.15	0.0346823603754\\
1.2	0.0237524791945\\
1.25	0.0178074958775\\
1.3	0.0118236244209\\
1.35	0.0080283446688\\
1.4	0.00629402354266\\
1.45	0.00455916608262\\
1.5	0.0031297518665\\
};
\addplot [color=green,line width=1.0pt,mark size=3pt,only marks,mark=+,mark options={solid}]
  table[row sep=crcr]{0.05	2.96162620282\\
0.1	3.55309359049\\
0.15	3.82749062353\\
0.2	3.98610442843\\
0.25	4.08994367825\\
0.3	4.16362261297\\
0.35	4.21734190142\\
0.4	4.25913923449\\
0.45	4.29242374856\\
0.5	4.31817727804\\
0.55	4.340090072\\
0.6	4.35695912708\\
0.65	4.37106955144\\
0.7	4.38160855894\\
0.75	4.38699936781\\
0.8	4.36036609912\\
0.85	4.18845063214\\
0.9	3.19109018262\\
0.95	1.34955054721\\
1	0.643225304118\\
1.05	0.387857087604\\
1.1	0.273479019388\\
1.15	0.197786879877\\
1.2	0.146477518277\\
1.25	0.116516029671\\
1.3	0.0899753988162\\
1.35	0.0717130525602\\
1.4	0.0558157680832\\
1.45	0.0418399217312\\
1.5	0.0356899709143\\
};
\addplot [color=mycolor2,line width=1.0pt,mark size=3pt,only marks,mark=x,mark options={solid}]
  table[row sep=crcr]{0.05	3.41090669646\\
0.1	4.18895540728\\
0.15	4.56553169656\\
0.2	4.78833956297\\
0.25	4.93647991141\\
0.3	5.04228422786\\
0.35	5.12113794178\\
0.4	5.18177470072\\
0.45	5.23094952011\\
0.5	5.26913151927\\
0.55	5.30103447991\\
0.6	5.32789760339\\
0.65	5.34706084089\\
0.7	5.36360636703\\
0.75	5.36934904778\\
0.8	5.35311726236\\
0.85	4.95569221402\\
0.9	3.16511697671\\
0.95	1.69609671345\\
1	1.14429908919\\
1.05	0.854537733667\\
1.1	0.64837136921\\
1.15	0.502980754341\\
1.2	0.405946323194\\
1.25	0.335170800866\\
1.3	0.263773243217\\
1.35	0.222232888851\\
1.4	0.179799010927\\
1.45	0.145410660492\\
1.5	0.125003195494\\
};
\addplot [color=purple,line width=1.0pt,mark size=3pt,only marks,mark=star,mark options={solid}]
  table[row sep=crcr]{0.05	4.21624736264231\\
0.1	5.86499150647989\\
0.15	7.014503625684\\
0.2	7.88932224538969\\
0.25	8.60903049315033\\
0.3	9.22660361104944\\
0.35	9.77765057186861\\
0.4	10.2335854120663\\
0.45	10.6723842474596\\
0.5	10.9970240462544\\
0.55	11.302651863981\\
0.6	11.5960418627813\\
0.65	11.8369964945988\\
0.7	11.9604065756021\\
0.75	12.356734802404\\
0.8	12.4786253974577\\
0.85	12.7243576101292\\
0.9	12.801997657958\\
0.95	12.9242520821987\\
1	13.0695773501555\\
1.05	13.1638030540024\\
1.1	13.3385669995014\\
1.15	13.4031255692071\\
1.2	13.5762476539716\\
1.25	13.74312939703\\
1.3	13.8491284419755\\
1.35	13.9969798799059\\
1.4	14.1660324355056\\
1.45	14.2076573505223\\
1.5	14.3885088513516\\
};
\addplot [color=gray,line width=1.0pt,mark size=3pt,only marks,mark=otimes,mark options={solid}]
  table[row sep=crcr]{0.05	6.16079592426522\\
0.1	8.46899322556305\\
0.15	9.7854223383504\\
0.2	10.5996203661161\\
0.25	11.6504608309166\\
0.3	12.205254081009\\
0.35	12.6178210631815\\
0.4	12.9463736515115\\
0.45	13.4155686420521\\
0.5	13.6965112238624\\
0.55	13.8798367643435\\
0.6	14.0718702541206\\
0.65	14.1977896196736\\
0.7	14.4751654597299\\
0.75	14.5908228565115\\
0.8	14.6824258593955\\
0.85	14.7141140370806\\
0.9	14.8462213454228\\
0.95	14.8125105901733\\
1	14.892501864892\\
1.05	15.0294743770988\\
1.1	14.9884528442042\\
1.15	15.0765917897296\\
1.2	15.1154506460725\\
1.25	15.1420701470676\\
1.3	15.2156477027293\\
1.35	15.2399400630622\\
1.4	15.3412276843842\\
1.45	15.345078160729\\
1.5	15.4501027483879\\
};
\addplot [color=yellow,line width=1.0pt,mark size=3pt,only marks,mark=square,mark options={solid}]
  table[row sep=crcr]{0.05	5.80608126912141\\
0.1	7.94503676548431\\
0.15	9.18776887510941\\
0.2	10.276418032343\\
0.25	10.979563822846\\
0.3	11.5779956453601\\
0.35	12.0615492947772\\
0.4	12.4117382007454\\
0.45	12.7614067698233\\
0.5	13.0515669074531\\
0.55	13.3180241807704\\
0.6	13.5274153643387\\
0.65	13.7213425217562\\
0.7	13.9062230307453\\
0.75	14.0411552931877\\
0.8	14.1511849684876\\
0.85	14.2560182735406\\
0.9	14.315264854472\\
0.95	14.3417640571818\\
1	14.4177135084414\\
1.05	14.4893385132445\\
1.1	14.4983608052703\\
1.15	14.5844387950132\\
1.2	14.7197828803039\\
1.25	14.7376093557155\\
1.3	14.8372770077123\\
1.35	14.8691787595096\\
1.4	14.9549885351378\\
1.45	15.00673002297\\
1.5	15.04502156101\\
};
\end{axis}
\end{tikzpicture}%
\caption{Spectral Efficiency, $\gamma$, of RS-IRSA and IRSA with ideal soliton ($L^{(1)}(x)$) with parameter $M$ (number of slots), modified soliton ($L^{(2)}(x)$) with parameter 10 and $L^{(3)}(x)$ as left degree distributions. $K=300, \frac{\tilde{E}_{s}}{N_{o}}=0.0009$.}
\label{fig:effsk3h}
\end{figure}
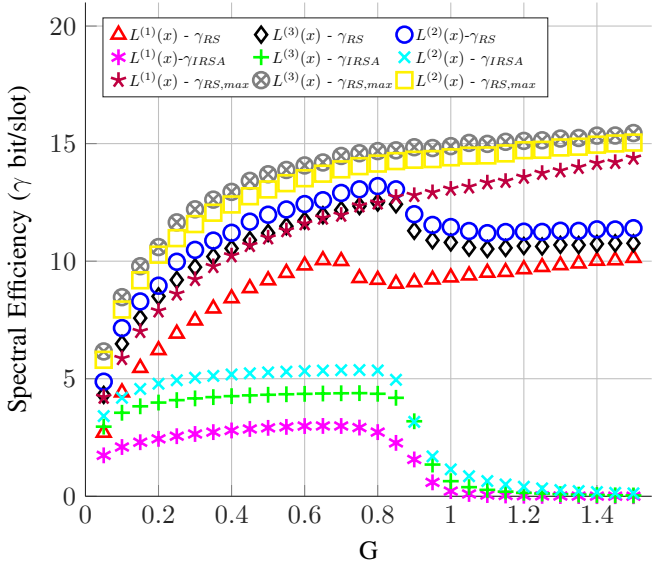

Figure~\ref{fig:rtthrok3h} shows that our scheme works even when $G$ exceeds one, with a throughput very close to $G$.
This implies that the receiver can decode almost all the messages, while preserving good efficiency.
Figures~\ref{fig:effck3h}~and~\ref{fig:effsk3h} show that both efficiency and spectral efficiency improve as $G$ increases, reaching their peak value around $G=0.8$.
This may be a good target operating point for $G = K/M$.

More importantly, our proposed rate adaptation scheme, RS-IRSA, uniformly outperforms IRSA, with very significant gains.
This holds true for the three expository distributions and different values of $G = K/M$.
This points to the need to carefully manage radio resources in practical implementations, and to leverage existing techniques such as MRC within the iterative decoding process.

\subsection{Power Adaptation}

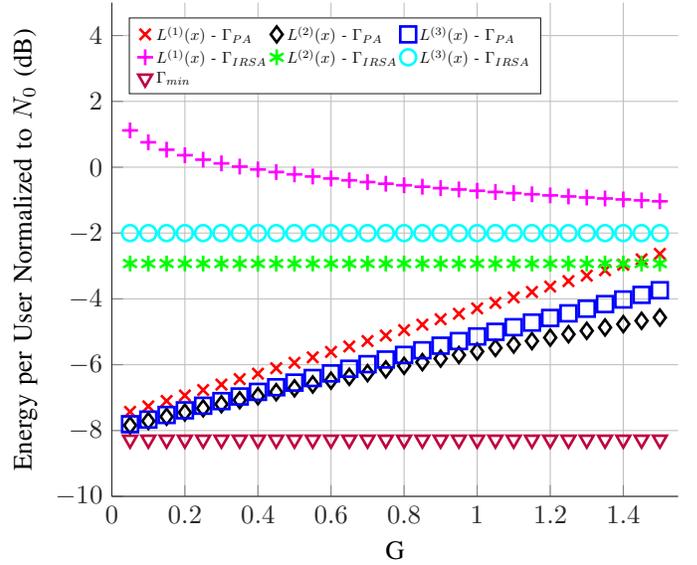
\begin{figure}[tbh]
%
%
%
\definecolor{mycolor1}{rgb}{1.00000,0.00000,1.00000}%
\definecolor{mycolor2}{rgb}{0.00000,1.00000,1.00000}%

\begin{tikzpicture}

\begin{axis}[%
width=0.85\columnwidth,
height=0.7417\columnwidth,
scale only axis,
every outer x axis line/.append style={white!15!black},
every x tick label/.append style={font=\color{white!15!black}},
xmin=0,
xmax=1.55,
xlabel={G},
xmajorgrids,
every outer y axis line/.append style={white!15!black},
every y tick label/.append style={font=\color{white!15!black}},
ymin=-10,
ymax=5,
ylabel={Energy per User Normalized to $N_{0}$ (dB)},
ymajorgrids,
axis x line*=bottom,
axis y line*=left,
legend columns=3,
legend entries={$L^{(1)}(x)$ - $\Gamma_{PA}$, $L^{(2)}(x)$ - $\Gamma_{PA}$, $L^{(3)}(x)$ - $\Gamma_{PA}$, $L^{(1)}(x)$ - $\Gamma_{IRSA}$, $L^{(2)}(x)$ - $\Gamma_{IRSA}$, $L^{(3)}(x)$ - $\Gamma_{IRSA}$, $\Gamma_{min}$},
legend style={at={(0.03,0.97)},anchor=north west,draw=white!15!black,fill=white,legend cell align=left,nodes={scale=0.6, transform shape}}
]


\addplot [color=red,line width=1.0pt,mark size=3.0pt,only marks,mark=x,mark options={solid}]
  table[row sep=crcr]{0.05	-7.43055\\
0.1	-7.2651\\
0.15	-7.09965\\
0.2	-6.9342\\
0.25	-6.76875\\
0.3	-6.6033\\
0.35	-6.43785\\
0.4	-6.2724\\
0.45	-6.10695\\
0.5	-5.9415\\
0.55	-5.77605\\
0.6	-5.6106\\
0.65	-5.44515\\
0.7	-5.2797\\
0.75	-5.11425\\
0.8	-4.9488\\
0.85	-4.78335\\
0.9	-4.6179\\
0.95	-4.45245\\
1	-4.287\\
1.05	-4.12155\\
1.1	-3.9561\\
1.15	-3.79065\\
1.2	-3.6252\\
1.25	-3.45975\\
1.3	-3.2943\\
1.35	-3.12885\\
1.4	-2.9634\\
1.45	-2.79795\\
1.5	-2.6325\\
};
\addplot [color=black,line width=1.0pt,mark size=3pt,only marks,mark=diamond,mark options={solid}]
  table[row sep=crcr]{0.05	-7.842189341\\
0.1	-7.709686196\\
0.15	-7.579261631\\
0.2	-7.450809634\\
0.25	-7.324230835\\
0.3	-7.199431928\\
0.35	-7.076325153\\
0.4	-6.954827839\\
0.45	-6.834861984\\
0.5	-6.716353881\\
0.55	-6.599233775\\
0.6	-6.483435558\\
0.65	-6.368896486\\
0.7	-6.255556926\\
0.75	-6.143360124\\
0.8	-6.03225199\\
0.85	-5.922180905\\
0.9	-5.813097545\\
0.95	-5.70495471\\
1	-5.597707183\\
1.05	-5.491311581\\
1.1	-5.385726234\\
1.15	-5.28091106\\
1.2	-5.17682746\\
1.25	-5.073438212\\
1.3	-4.970707376\\
1.35	-4.868600205\\
1.4	-4.767083066\\
1.45	-4.666123354\\
1.5	-4.565689427\\
};
\addplot [color=blue,line width=1.0pt,mark size=3pt,only marks,mark=square,mark options={solid}]
  table[row sep=crcr]{0.05	-7.80765\\
0.1	-7.6673\\
0.15	-7.52695\\
0.2	-7.3866\\
0.25	-7.24625\\
0.3	-7.1059\\
0.35	-6.96555\\
0.4	-6.8252\\
0.45	-6.68485\\
0.5	-6.5445\\
0.55	-6.40415\\
0.6	-6.2638\\
0.65	-6.12345\\
0.7	-5.9831\\
0.75	-5.84275\\
0.8	-5.7024\\
0.85	-5.56205\\
0.9	-5.4217\\
0.95	-5.28135\\
1	-5.141\\
1.05	-5.00065\\
1.1	-4.8603\\
1.15	-4.71995\\
1.2	-4.5796\\
1.25	-4.43925\\
1.3	-4.2989\\
1.35	-4.15855\\
1.4	-4.0182\\
1.45	-3.87785\\
1.5	-3.7375\\
};

\addplot [color=mycolor1,line width=1.0pt,mark size=3.0pt,only marks,mark=+,mark options={solid}]
  table[row sep=crcr]{0.05	1.1180119279035\\
0.1	0.757416885180807\\
0.15	0.531713234627515\\
0.2	0.364148253610965\\
0.25	0.229571057812042\\
0.3	0.116431075048859\\
0.35	0.0184179464753026\\
0.4	-0.0683104390408911\\
0.45	-0.146274895179797\\
0.5	-0.217222609539748\\
0.55	-0.28241661246109\\
0.6	-0.342800892694999\\
0.65	-0.399100357763214\\
0.7	-0.451884317356267\\
0.75	-0.501608420751609\\
0.8	-0.5486432907565\\
0.85	-0.593294631206892\\
0.9	-0.635817692779049\\
0.95	-0.67642790161985\\
1	-0.715308814578546\\
1.05	-0.752618171948408\\
1.1	-0.788492570591876\\
1.15	-0.823051119630475\\
1.2	-0.856398334346822\\
1.25	-0.888626451840606\\
1.3	-0.919817302256777\\
1.35	-0.950043834528538\\
1.4	-0.979371370735112\\
1.45	-1.00785864522599\\
1.5	-1.03555867152635\\
};

\addplot [color=green,line width=1.0pt,mark size=3.0pt,only marks,mark=asterisk,mark options={solid}]
  table[row sep=crcr]{0.05	-2.92530371556497\\
0.1	-2.92530371556497\\
0.15	-2.92530371556497\\
0.2	-2.92530371556497\\
0.25	-2.92530371556497\\
0.3	-2.92530371556497\\
0.35	-2.92530371556497\\
0.4	-2.92530371556497\\
0.45	-2.92530371556497\\
0.5	-2.92530371556497\\
0.55	-2.92530371556497\\
0.6	-2.92530371556497\\
0.65	-2.92530371556497\\
0.7	-2.92530371556497\\
0.75	-2.92530371556497\\
0.8	-2.92530371556497\\
0.85	-2.92530371556497\\
0.9	-2.92530371556497\\
0.95	-2.92530371556497\\
1	-2.92530371556497\\
1.05	-2.92530371556497\\
1.1	-2.92530371556497\\
1.15	-2.92530371556497\\
1.2	-2.92530371556497\\
1.25	-2.92530371556497\\
1.3	-2.92530371556497\\
1.35	-2.92530371556497\\
1.4	-2.92530371556497\\
1.45	-2.92530371556497\\
1.5	-2.92530371556497\\
};
\addplot [color=mycolor2,line width=1.0pt,mark size=3.0pt,only marks,mark=o,mark options={solid}]
  table[row sep=crcr]{0.05	-2.00194843403453\\
0.1	-2.00194843403453\\
0.15	-2.00194843403453\\
0.2	-2.00194843403453\\
0.25	-2.00194843403453\\
0.3	-2.00194843403453\\
0.35	-2.00194843403453\\
0.4	-2.00194843403453\\
0.45	-2.00194843403453\\
0.5	-2.00194843403453\\
0.55	-2.00194843403453\\
0.6	-2.00194843403453\\
0.65	-2.00194843403453\\
0.7	-2.00194843403453\\
0.75	-2.00194843403453\\
0.8	-2.00194843403453\\
0.85	-2.00194843403453\\
0.9	-2.00194843403453\\
0.95	-2.00194843403453\\
1	-2.00194843403453\\
1.05	-2.00194843403453\\
1.1	-2.00194843403453\\
1.15	-2.00194843403453\\
1.2	-2.00194843403453\\
1.25	-2.00194843403453\\
1.3	-2.00194843403453\\
1.35	-2.00194843403453\\
1.4	-2.00194843403453\\
1.45	-2.00194843403453\\
1.5	-2.00194843403453\\
};
\addplot [color=purple,line width=1.0pt,mark size=3pt,only marks,mark=triangle,mark options={solid,rotate=180}]
  table[row sep=crcr]{0.05	-8.2769383591424\\
0.1	-8.2769383591424\\
0.15	-8.2769383591424\\
0.2	-8.2769383591424\\
0.25	-8.2769383591424\\
0.3	-8.2769383591424\\
0.35	-8.2769383591424\\
0.4	-8.2769383591424\\
0.45	-8.2769383591424\\
0.5	-8.2769383591424\\
0.55	-8.2769383591424\\
0.6	-8.2769383591424\\
0.65	-8.2769383591424\\
0.7	-8.2769383591424\\
0.75	-8.2769383591424\\
0.8	-8.2769383591424\\
0.85	-8.2769383591424\\
0.9	-8.2769383591424\\
0.95	-8.2769383591424\\
1	-8.2769383591424\\
1.05	-8.2769383591424\\
1.1	-8.2769383591424\\
1.15	-8.2769383591424\\
1.2	-8.2769383591424\\
1.25	-8.2769383591424\\
1.3	-8.2769383591424\\
1.35	-8.2769383591424\\
1.4	-8.2769383591424\\
1.45	-8.2769383591424\\
1.5	-8.2769383591424\\
};

\end{axis}
\end{tikzpicture}%
\caption{Energy per uses normalized to $N_{0}$ of PA-IRSA and IRSA with ideal soliton ($L^{(1)}(x)$) with parameter $M$ (number of slots), modified soliton ($L^{(2)}(x)$) with parameter 10 and $L^{(3)}(x)$ as left degree distributions. $K=300, \hat{R}=10, \mathfrak{L}=100$.}
\label{fig:powk3h}
\end{figure}

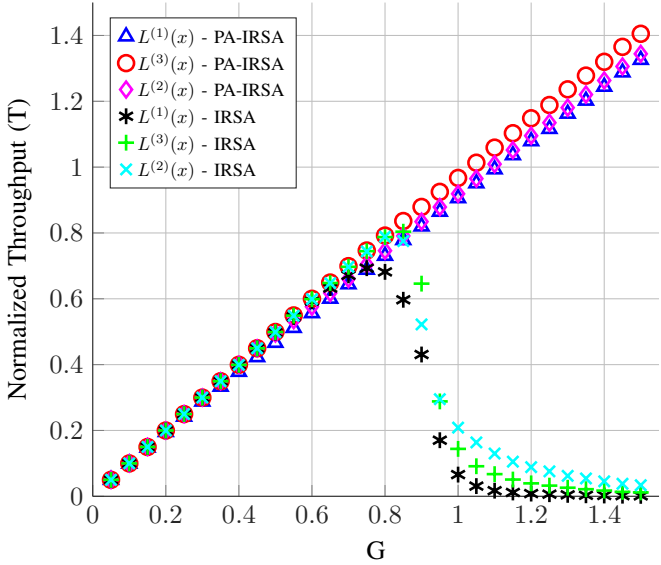
\begin{figure}[tbh]
%
%
%
\definecolor{mycolor1}{rgb}{1.00000,0.00000,1.00000}%
\definecolor{mycolor2}{rgb}{0.00000,1.00000,1.00000}%
\begin{tikzpicture}

\begin{axis}[%
width=0.85\columnwidth,
height=0.7417\columnwidth,
scale only axis,
every outer x axis line/.append style={white!15!black},
every x tick label/.append style={font=\color{white!15!black}},
xmin=0,
xmax=1.55,
xlabel={G},
xmajorgrids,
every outer y axis line/.append style={white!15!black},
every y tick label/.append style={font=\color{white!15!black}},
ymin=0,
ymax=1.5,
ylabel={Normalized Throughput (T)},
ymajorgrids,
axis x line*=bottom,
axis y line*=left,
legend entries={$L^{(1)}(x)$ - PA-IRSA, $L^{(3)}(x)$ - PA-IRSA, $L^{(2)}(x)$ - PA-IRSA, $L^{(1)}(x)$ - IRSA, $L^{(3)}(x)$ - IRSA, $L^{(2)}(x)$ - IRSA},
legend style={at={(0.03,0.97)},anchor=north west,draw=white!15!black,fill=white,legend cell align=left,nodes={scale=0.7, transform shape}}
]
\addplot [color=blue,line width=1.0pt,mark size=3.0pt,only marks,mark=triangle,mark options={solid}]
  table[row sep=crcr]{0.05	0.049847359\\
0.1	0.099019327\\
0.15	0.14778\\
0.2	0.195225183\\
0.25	0.241873439\\
0.3	0.287726\\
0.35	0.332981352\\
0.4	0.378110519\\
0.45	0.422710645\\
0.5	0.466294509\\
0.55	0.51147619\\
0.6	0.556308\\
0.65	0.600359307\\
0.7	0.644088578\\
0.75	0.688221945\\
0.8	0.730571809\\
0.85	0.778303116\\
0.9	0.819796407\\
0.95	0.86443038\\
1	0.905644518\\
1.05	0.95084965\\
1.1	0.993285714\\
1.15	1.036218391\\
1.2	1.078704\\
1.25	1.116103734\\
1.3	1.162415584\\
1.35	1.202013453\\
1.4	1.244283721\\
1.45	1.287797101\\
1.5	1.325860697\\
};
\addplot [color=red,line width=1.0pt,mark size=3pt,only marks,mark=o,mark options={solid}]
  table[row sep=crcr]{0.05	0.049935844\\
0.1	0.099812396\\
0.15	0.149731\\
0.2	0.199750833\\
0.25	0.249696087\\
0.3	0.299859\\
0.35	0.349574592\\
0.4	0.399435419\\
0.45	0.449704648\\
0.5	0.499144759\\
0.55	0.549397436\\
0.6	0.599974\\
0.65	0.649309524\\
0.7	0.699137529\\
0.75	0.746623441\\
0.8	0.792007979\\
0.85	0.836249292\\
0.9	0.879023952\\
0.95	0.92485443\\
1	0.967215947\\
1.05	1.013398601\\
1.1	1.059216117\\
1.15	1.103042146\\
1.2	1.148424\\
1.25	1.188726141\\
1.3	1.236324675\\
1.35	1.27806278\\
1.4	1.32\\
1.45	1.365376812\\
1.5	1.405099502\\
};
\addplot [color=mycolor1,line width=1.0pt,mark size=3pt,only marks,mark=diamond,mark options={solid}]
  table[row sep=crcr]{0.05	0.049955341\\
0.1	0.099890037\\
0.15	0.1499285\\
0.2	0.199831446\\
0.25	0.249730225\\
0.3	0.299969\\
0.35	0.349609557\\
0.4	0.399395473\\
0.45	0.449443778\\
0.5	0.494682196\\
0.55	0.535\\
0.6	0.575934\\
0.65	0.617575758\\
0.7	0.660410256\\
0.75	0.703972569\\
0.8	0.74612766\\
0.85	0.79107932\\
0.9	0.834505988\\
0.95	0.878702532\\
1	0.918853821\\
1.05	0.965111888\\
1.1	1.009098901\\
1.15	1.051697318\\
1.2	1.095312\\
1.25	1.134850622\\
1.3	1.180705628\\
1.35	1.220524664\\
1.4	1.26324186\\
1.45	1.305835749\\
1.5	1.344109453\\
};
\addplot [color=black,line width=1.0pt,mark size=3pt,only marks,mark=asterisk,mark options={solid}]
  table[row sep=crcr]{0.05	0.0499916680553\\
0.1	0.0999660113296\\
0.15	0.149991\\
0.2	0.199850099933\\
0.25	0.249752706078\\
0.3	0.299942\\
0.35	0.349548951049\\
0.4	0.399287616511\\
0.45	0.449452773613\\
0.5	0.497216306156\\
0.55	0.545018315018\\
0.6	0.592544\\
0.65	0.632569264069\\
0.7	0.671221445221\\
0.75	0.692992518703\\
0.8	0.681736702128\\
0.85	0.597\\
0.9	0.43074251497\\
0.95	0.170329113924\\
1	0.0659102990033\\
1.05	0.030951048951\\
1.1	0.0170769230769\\
1.15	0.0116283524904\\
1.2	0.008244\\
1.25	0.00636514522822\\
1.3	0.00437229437229\\
1.35	0.00305381165919\\
1.4	0.00246511627907\\
1.45	0.00184057971014\\
1.5	0.00129353233831\\
};
\addplot [color=green,line width=1.0pt,mark size=3pt,only marks,mark=+,mark options={solid}]
  table[row sep=crcr]{0.05	0.0499913347775\\
0.1	0.0999623458847\\
0.15	0.149994\\
0.2	0.199850099933\\
0.25	0.249755203997\\
0.3	0.299944\\
0.35	0.349534965035\\
0.4	0.399299600533\\
0.45	0.44952173913\\
0.5	0.498727121464\\
0.55	0.548833333333\\
0.6	0.598972\\
0.65	0.647924242424\\
0.7	0.697179487179\\
0.75	0.744713216958\\
0.8	0.787449468085\\
0.85	0.803841359773\\
0.9	0.646038922156\\
0.95	0.288259493671\\
1	0.144019933555\\
1.05	0.0912587412587\\
1.1	0.0673223443223\\
1.15	0.0508659003831\\
1.2	0.039284\\
1.25	0.0323858921162\\
1.3	0.0260649350649\\
1.35	0.0215022421525\\
1.4	0.0173441860465\\
1.45	0.0134927536232\\
1.5	0.0118457711443\\
};
\addplot [color=mycolor2,line width=1.0pt,mark size=3pt,only marks,mark=x,mark options={solid}]
  table[row sep=crcr]{0.05	0.049990668222\\
0.1	0.099965344885\\
0.15	0.1499965\\
0.2	0.199848767488\\
0.25	0.249751873439\\
0.3	0.299914\\
0.35	0.349546620047\\
0.4	0.399271637816\\
0.45	0.449503748126\\
0.5	0.498695507488\\
0.55	0.54871978022\\
0.6	0.598982\\
0.65	0.647649350649\\
0.7	0.696869463869\\
0.75	0.743815461347\\
0.8	0.788484042553\\
0.85	0.77533427762\\
0.9	0.522149700599\\
0.95	0.295091772152\\
1	0.208624584718\\
1.05	0.163664335664\\
1.1	0.129882783883\\
1.15	0.105233716475\\
1.2	0.088548\\
1.25	0.075755186722\\
1.3	0.0621212121212\\
1.35	0.0541614349776\\
1.4	0.0454046511628\\
1.45	0.0381014492754\\
1.5	0.0337064676617\\
};
\end{axis}
\end{tikzpicture}%
\caption{Normalized throughput, T, of PA-IRSA and IRSA with ideal soliton ($L^{(1)}(x)$) with parameter $M$ (number of slots), modified soliton ($L^{(2)}(x)$) with parameter 10 and $L^{(3)}(x)$ as left degree distributions. $K=300, \hat{R}=10, \mathfrak{L}=100$.}
\label{fig:powthrok3h}
\end{figure}

Given our focus on system operation in the low SNR regime, we fix rate $\hat{R}$ at 10 and $\mathfrak{L}=100$.
The best value for $\mu$ and the repetition distribution are optimized numerically such that receiver can decode at least 90\% of the messages for $K=300$.
Figure~\ref{fig:powk3h} plots the energy per user normalized to $N_{0}$.
Figure~\ref{fig:powthrok3h} shows a comparison between the normalized throughputs for PA-IRSA and IRSA.
The results are averaged over 1000 independent trials for every operating point.
As $G$ grows, the throughput level is maintained by increasing energy.
This is necessary because the number of interferers in each slot also grows.
Consequently, at larger values for $G = K/M$, performance gains are not as significant and $\Gamma_{\PA}$ converges to $\Gamma_{\IRSA}$.
The ideal soliton features the largest marginal gain.

Much like in the rate selection case, the modified soliton produces the best performance among these three distributions.
In addition to significant power gains, the modified soliton also shows better normalized throughput.
Around $G=0.8$, where IRSA achieves its best normalized throughput, PA-IRSA features comparable throughput; yet every device, on average, employs at least 3~dB less energy than IRSA.

Our proposed scheme, PA-IRSA, works even when $G$ exceeds one.
For instance, at $G=1.5$, the normalized throughput corresponding to the modified soliton is 1.4, and it uses 1.5~dB less energy than IRSA.
We note that the throughput of IRSA drops rapidly for $G>0.8$, and it becomes zero at $G=1.5$.
The data points for IRSA in Fig.~\ref{fig:powk3h} when $G>0.8$ are present solely to ensure an easy comparison between IRSA and PA-IRSA.

In the case of power adaptation, again, the most important lesson is to manage radio resources carefully and to take advantage of MRC along with successive interference cancellation.
Maximum ratio combining is an optimal means to go beyond single-slot decoding, and it yields significant gains in several instances.
Once this algorithmic enhancement is incorporated into the system, this opens up new possibilities for power adaptation.

\subsection{RS and PA Comparison}

\begin{figure}[t]
%
%
%
%
%

\definecolor{mycolor1}{rgb}{1.00000,0.00000,1.00000}%
\definecolor{mycolor2}{rgb}{0.00000,1.00000,1.00000}%
\begin{tikzpicture}

\begin{axis}[%
width=0.85\columnwidth,
height=0.7417\columnwidth,
scale only axis,
every outer x axis line/.append style={white!15!black},
every x tick label/.append style={font=\color{white!15!black}},
xmin=2,
xmax=18,
xlabel={Average Rate of Users (bit/sec)},
xmajorgrids,
every outer y axis line/.append style={white!15!black},
every y tick label/.append style={font=\color{white!15!black}},
ymin=-15,
ymax=8,
ylabel={Energy per User Normalized to $N_{0}$ (dB)},
ymajorgrids,
axis x line*=bottom,
axis y line*=left,
legend columns=3,
legend entries={$L^{(1)}(x)$ - IRSA, $L^{(1)}(x)$ - PA-IRSA, $L^{(1)}(x)$ - RS-IRSA, $L^{(2)}(x)$ - IRSA, $L^{(2)}(x)$ - PA-IRSA, $L^{(2)}(x)$ - RS-IRSA, $L^{(3)}(x)$ - IRSA, $L^{(3)}(x)$ - PA-IRSA, $L^{(3)}(x)$ - RS-IRSA},
legend style={at={(0.03,0.97)},anchor=north west,draw=white!15!black,fill=white,legend cell align=left,nodes={scale=0.6, transform shape}}
]

\addplot [color=red,dotted,line width=1.0pt,mark size=3.0pt,mark=+,mark options={solid}]
  table[row sep=crcr]{4	-4.71158204565415\\
7	-2.18974506002762\\
10	-0.548643290756499\\
13	0.683497765978696\\
16	1.67859655323148\\
};
\addplot [color=black,dash pattern=on 1pt off 3pt on 3pt off 3pt,line width=1.0pt,mark size=3.0pt,mark=x,mark options={solid}]
  table[row sep=crcr]{3.343998809	-12.1384911896381\\
4.776476393	-9.9945766620735\\
7.009133821	-7.58362867537836\\
9.060458782	-5.34259299592629\\
11.51078453	-3.13948304587684\\
14.44720639	0.395508308867768\\
};
\addplot [color=blue,dashed,line width=1pt,mark size=3pt,mark=diamond,mark options={solid}]
  table[row sep=crcr]{3.34399880894	-12.7073547472141\\
4.77647639284	-10.7074510621141\\
7.00913382109	-8.7073620128141\\
9.06045878204	-6.7073538203141\\
11.5107845326	-4.7076474127141\\
14.24543018	-2.7887921516141\\
};
\addplot [color=mycolor1,dotted,line width=1.0pt,mark size=3pt,mark=asterisk,mark options={solid}]
  table[row sep=crcr]{4	-7.08824247046262\\
7	-4.56640548483609\\
10	-2.92530371556497\\
13	-1.69316265882977\\
16	-0.698063871576993\\
};
\addplot [color=green,dash pattern=on 1pt off 3pt on 3pt off 3pt,line width=1.0pt,mark size=3pt,mark=triangle,mark options={solid,rotate=180}]
  table[row sep=crcr]{3.563760844	-12.2343809205871\\
5.408317132	-9.94251136776014\\
8.105023806	-7.5155398160679\\
11.8999997	-4.61722979779478\\
16.49289477	-1.42086242447589\\
};
\addplot [color=mycolor2,dash pattern=on 1pt off 3pt on 3pt off 3pt,line width=1.0pt,mark size=3.0pt,mark=star,mark options={solid}]
  table[row sep=crcr]{3.56376084433	-12.7054037907226\\
5.40831713217	-10.7055001056226\\
8.10502380604	-8.70541105632257\\
11.8999997033	-6.70540286382257\\
16.4928947721	-4.70569645632257\\
};
\addplot [color=purple,dotted,line width=1.0pt,mark size=3pt,mark=o,mark options={solid}]
  table[row sep=crcr]{4	-6.16488718893218\\
7	-3.64305020330565\\
10	-2.00194843403453\\
13	-0.769807377299334\\
16	0.225291409953447\\
};
\addplot [color=yellow,dash pattern=on 1pt off 3pt on 3pt off 3pt,line width=1.0pt,mark size=3pt,mark=square,mark options={solid}]
  table[row sep=crcr]{3.54	-11.72\\
5.382	-10.054\\
8.011	-7.103\\
11.63	-4.042\\
15.66	-1.2444\\
};
\addplot [color=gray,dash pattern=on 1pt off 3pt on 3pt off 3pt,line width=1.0pt,mark size=3pt,mark=triangle,mark options={solid}]
  table[row sep=crcr]{3.53970505937	-12.7054037907921\\
5.38150063822	-10.7055001056921\\
8.01129019053	-8.70541105639213\\
11.6254584612	-6.70540286389213\\
15.6572999651	-4.70569645629213\\
};
\end{axis}
\end{tikzpicture}%
\caption{RS-IRSA, PA-IRSA and IRSA comparison with ideal soliton ($L^{(1)}(x)$) with parameter $M$ (number of slots), modified soliton ($L^{(2)}(x)$) with parameter 10 and $L^{(3)}(x)$ as left degree distributions. $K=300, G=0.8, \mathfrak{L}=100$.}
\label{fig:rspacmpr}
\end{figure}
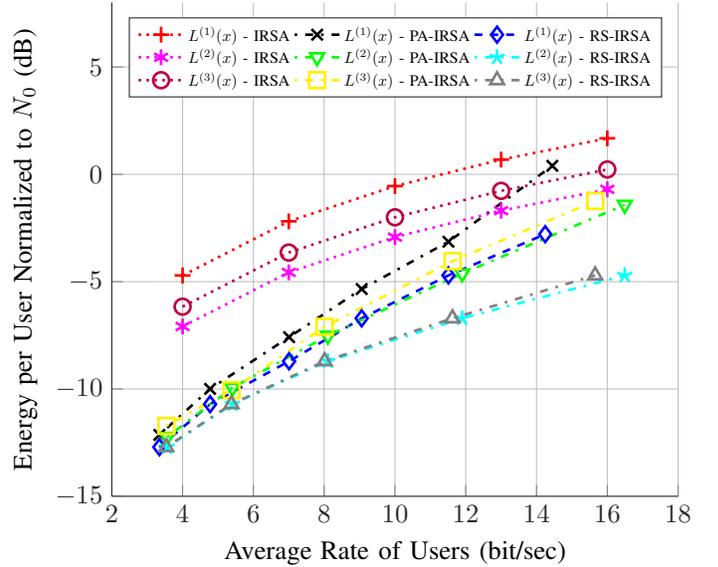

We compare alternate implementations under system parameter $G=0.8$ because all the schemes, IRSA, PA-IRSA, and RS-IRSA, reach their best performance around this operating point.
Figure~\ref{fig:rspacmpr} displays the respective performance of these schemes.
Various values of $E_{s}/N_{0}$ are fed to the RS-IRSA scheme to find the best average rate per user, $\overline{R}_{\RS}$, such that normalized throughput is at least 0.78.
Energy levels are chosen as to remain in the low SNR regime.
The $y$-axis for RS-IRSA corresponds to $l_{\avg}E_{s}/N_{0}$.
The optimal rates for rate adaptation, $\overline{R}_{\RS}$, are selected as input to the PA-IRSA scheme.
Then, we find the minimum energy levels, $\Gamma_{\PA}$, such that the normalized throughputs are at least 0.78 at the target rates.
Finally, since IRSA only uses one replica during the decode a message, the energy per channel use for IRSA is derived directly from the definition of capacity.
It is subsequently multiplied by $l_{\avg}$ to form a data point.
The results are averaged over 1000 independent trials for every sample.

As the rate increases, the energy level also climbs.
This produces higher interference levels, which leads to lower performances for RS-IRSA and PA-IRSA.
Consequently, the gains are not as pronounced at higher rates.
RS-IRSA features better performance than PA-IRSA.
This can be explained by the fact that, in PA-IRSA, two variables are estimated, i.e., number of interferers in a slot and their power levels; whereas, only the number of interferers is estimated in RS-IRSA.
At low rates or, equivalently, very low SNRs, PA-IRSA and RS-IRSA show similar performance.
As anticipated, the modified soliton produces the best performance among the candidate distributions.

\section{Conclusion}
In this paper, we introduce two new random access protocols for the slotted GMAC channel.
Our proposed schemes, RS-IRSA and PA-IRSA, employ estimates for the maximum achievable rate per user and the maximum achievable SINR per user, respectively, to adjust code rates or transmit powers.
The estimation process takes into account the repetition profile of every device, and the fact that MRC and SIC are utilized at the receiver.
These schemes allow users with higher degrees to increase their rates or lower their transmit powers, taking advantage of their effective interference pattern. 

Also, we show that, with a finite number of users and in low SNR regime, the sum rate for all the users can reach 80\% of the GMAC capacity with coordination.
This, we feel, is admirable given the information asymmetry between the two problem formulations.
In this same setting, RS-IRSA showcases a 50\% improvement over the state-of-the-art in slotted random multiple access. 
Moreover, in this regime, PA-IRSA yields a minimum gain of 3~dB in transmit power with respect to IRSA.
Altogether, our contributions demonstrate that MRC and SIC should both be employed at the receiver.
Furthermore, once these enhancements are in place, it is valuable to tailor the left degree distribution to this reality.
This results in important gains in terms of rates or power savings, especially in the low SNR regime.
These benefits hinge on the ability to accurately predict effective interference levels experienced during the decoding process.


There are several possible avenues for future inquiries.
While we introduce a new viewpoint for random access and show excellent performance improvements, we do not present an optimal family of distributions for slot selection.
This optimization problem appears very challenging, as it relies on a multitude of factors including $K$, $M$, and $N_{0}$.
Still, providing better insights for the selection of good distributions would be valuable.
The framework studied in this paper focuses on a single frame.
Expanding our findings to scenarios with multiple frames or with frame length adaptation forms an interesting challenge.
Finally, while the results are very promising, many practical concerns should be address before these techniques are pushed into standards.
This too warrants attention.



\bibliographystyle{ieeetr}
\bibliography{IEEEabrv,MACcollision}

\end{document}